\documentclass[reprint,amssymb, amsmath, aip,cha,twocolumn]{revtex4-1}
\usepackage{graphicx}
\usepackage[caption = false]{subfig}
\usepackage{color}
\usepackage{epsfig}
\usepackage{ifpdf}
\usepackage{url}%
\usepackage{bm}
\usepackage{soul}
\usepackage{natbib}
\usepackage[colorlinks=true,linkcolor=blue]{hyperref}%
\usepackage{cleveref}
\usepackage{soul}
\expandafter\ifx\csname package@font\endcsname\relax\else
 \expandafter\expandafter
 \expandafter\usepackage
 \expandafter\expandafter
 \expandafter{\csname package@font\endcsname}%
\fi
\begin{document}
\title{DPEx-II: A New Dusty Plasma Device Capable of Producing Large Sized DC Coulomb Crystals}%

\author{Saravanan Arumugam}
\affiliation{Institute For Plasma Research, Bhat, Gandhinagar-382428, India.}%
\email{sarvaanphysics@gmail.com}
\author{P. Bandyopadhyay}
\affiliation{Institute For Plasma Research, Bhat, Gandhinagar-382428, India.}%
\affiliation{Homi Bhabha National Institute, Training School Complex, Anushakti Nagar, Mumbai-400085 India.}
\author{Swarnima Singh}
\affiliation{Institute For Plasma Research, Bhat, Gandhinagar-382428, India.}%
\affiliation{Homi Bhabha National Institute, Training School Complex, Anushakti Nagar, Mumbai-400085 India.}
\author{M. G. Hariprasad}
\affiliation{Institute For Plasma Research, Bhat, Gandhinagar-382428, India.}%
\affiliation{Homi Bhabha National Institute, Training School Complex, Anushakti Nagar, Mumbai-400085 India.}
\author{Dinesh Rathod}
\affiliation{Department of Research and Development, University of Petroleum and Energy Studies, Dehradun, Uttarakhand-248007, India.}%
\author{Garima Arora}
\affiliation{Institute For Plasma Research, Bhat, Gandhinagar-382428, India.}%
\affiliation{Homi Bhabha National Institute, Training School Complex, Anushakti Nagar, Mumbai-400085 India.}
\author{A. Sen}
\affiliation{Institute For Plasma Research, Bhat, Gandhinagar-382428, India.}%
\affiliation{Homi Bhabha National Institute, Training School Complex, Anushakti Nagar, Mumbai-400085 India.}
\date{\today}
\begin{abstract}
The creation of a spatially extended stable DC complex plasma crystal is a big experimental challenge and a topical area of research in the field of dusty plasmas. In this paper we describe a newly built and commissioned dusty plasma experimental device, DPEx-II,  at the Institute for Plasma Research. {The device can support the formation of large sized Coulomb crystals in a DC glow discharge plasma.} The plasma in this L-shaped table-top glass chamber is produced between a circular anode and a long tray shaped cathode. It is characterized with the help of various electrostatic probes over a range of discharge conditions. The dust particles are  introduced by a dust dispenser to form a strongly coupled Coulomb crystal in the cathode sheath region. The unique asymmetric electrode configuration minimizes the heating of dust particles and facilitates the formation of crystalline structures with a maximum achievable dimension of  40~cm~$\times$~15~cm using this device.  A larger crystal has numerous advantages over smaller ones, such as higher structural homogeneity, fewer defects, lower statistical errors due to finite size effects \textit{etc}. A host of diagnostics tools are provided to characterize the Coulomb crystal.  Results of a  few initial experiments aimed at demonstrating the technical capabilities of the device and its potential for future dusty plasma research, are reported.
\end{abstract}
\maketitle
\section{Introduction}\label{sec:intro}
A dusty or complex plasma comprises of a conventional two component plasma embedded with micron or sub-micron sized solid particles \cite{collect1,collect2}. These solid particles get electrically charged in the plasma environment due to the bombardment of electrons and ions on them. Since electrons are more mobile than ions in a low-temperature laboratory plasma there is a higher flux of electrons than ions on the dust particles making them negatively charged. The amount of charge on a dust particle depends on the size of the dust particle as well as the ambient plasma parameters. The typical amount of charge on a dust particle in a  laboratory dusty plasma system lies in the range of $10^{3}$e to $10^{5}$e \cite{dustcharge1,dustcharge2} where $e$ is the electronic charge. The presence of these highly charged dust particles introduces additional degrees of freedom and a variety of time scales and length scales in the dusty plasma system. Consequently, a dusty plasma medium exhibits rich collective dynamics.  Another important characteristic of a dusty plasma is that the dust component can develop strong inter-particle correlations due to the high charge on each particle and its low thermal energy.  A quantitative measure of this interaction is given by the so called Coulomb parameter which is the ratio of the electrostatic potential energy to the thermal energy. When the Coulomb parameter becomes very large the dust component can go into a crystalline state. Dust crystals are novel objects that have attracted a lot of attention in recent times. Using a suitable arrangement of electrostatic force to counter the downward gravitational force acting on dust particles, one can form various configurations of 2-D \cite{dustcharge1,dustcharge2} and 3-D \cite{DC_3D,Polyakov2013} dust crystal in the laboratory. Depending on the discharge conditions, the dust system can pass through different phases like crystalline, hexatic, and liquid states \cite{Trans, Polyakov_2019}. The phased nature of the dusty plasma medium can be characterized by the Coulomb coupling parameter \cite{Hari1,coupling1}. \par
Most past experiments on dusty plasma crystals have been carried out in RF discharges \cite{RF1, RF2, RF3,void3}. There are only a few experiments on  the plasma crystal that have been done in  DC discharges \cite{void1,void2,DCdusty,Vasilyak,Balabanov}. It is found that  the plasma sustaining mechanism has a significant influence on the dynamics of dust particles. In the DC plasma, the ions get affected significantly in the cathode sheath region, whereas in RF plasmas, they cannot respond to the high frequency oscillating electric field. This makes a significant difference on the dust charging processes and heating mechanism of dust particles by ion streaming \cite{DCRF0, DC_3D, DCRF1, DCRF2, DCRF3}. Hence, the mode of discharges has a direct control over the Coulomb coupling parameter as it solely depends on the dust charge and the dust temperature. Therefore, the discharge window for a specific observation in dusty plasma crystal becomes completely different in DC and RF discharges \cite{DCRF2,DCRF3}. \par
The typical spatial extent of dust crystals in laboratory RF discharges are in the range  of a few cm$^2$. It is due to the fact that the particles levitate in the sheath of the bottom electrode in a parallel plate configuration having small plasma volume. The size of the crystal, with a smaller number of particles (a few tens to few hundreds) is good enough to study the dynamics of particles at an individual level \cite{finitedust1}. However, for exploring the thermodynamics of Coulomb crystal, the number of dust particles has to be as large as possible to minimize the error associated with the statistical fluctuation ($\propto\frac{1}{\sqrt{N}}$, where \textit{N} is the number of particles) \cite{Therfluc}. Moreover, in the measurement of any intensive physical quantity of the Coulomb crystal, one natural way of eliminating the boundary effect is to perform experiments with larger crystals. A dusty plasma medium with larger spatial extension is a natural prerequisite for observing wave phenomena in it. A dusty plasma crystal having few dust particles doesn't support the excitation and propagation of the wave and hence a dust crystal of reasonably large dimension is indispensable to accommodate few wavelengths of the excited wave \cite{garima1,soliton,surabhi_acoustic}. Additionally, studies like, void formation \cite{void1,void2,void3}, phase co-existence \cite{Phasecoex}, phase transition \cite{phasetrans,Trans}, effect of spatial inhomogeneity of dust density on the dynamics of dust particles \textit{etc.} cannot be investigated with few dust particles in the crystal \cite{dengrad1,dengrad2}.\par
Recently the first-ever observation of a plasma crystal in the cathode sheath of a DC glow discharge plasma was reported in  the DPEx device \cite{DPEX} at the Institute for Plasma Research.  However, the size of the crystal was limited due to some inherent constraints of the device.  The principal focus of DPEx device was to conduct experiments on the excitations of nonlinear waves in a flowing dusty plasma through a narrow channel. Consequently  there was very little room to form a  large size Coulomb crystal. Due to the small size of the dust crystal (less than a cm), there persists always an inhomogeneity in the dust density, temperature and Coulomb coupling parameter in its spatial extent \cite{Hari1}.  Moreover, the smaller confining boundaries lead to the formation of natural defects which are inevitable in a finite-sized crystal and the number of defects increases with the reduction in the crystal size. To address some of these challenges, we have built a new L-shaped  DC glow discharge dusty plasma experimental device, named DPEx-II, which has several inbuilt advantages for performing experiments with large crystals. Apart from permitting easy optical diagnostics access, the asymmetric electrode configuration with a large cathode area allows us to form a large dust crystal with various confining potential structures.  Compared to the DPEx device, the larger area of cathode serves to reduce the ion flux on the dust surface and maximizes the favourable condition to form a stable dust crystal over a wide range of dicharge conditions. This newly built device helps us to understand the underlying physics by performing various experiments which could not be performed in DPEx device \cite{DPEX}.\par
In this paper, we report the principle design, and operation of the newly developed DC glow discharge dusty plasma experimental device with a detailed characterization of plasma and dusty plasma crystal. We also present in  brief the ongoing experiments in this device. The paper is organized as follows: in Sec.~\ref{sec:setup}, the details of the experimental device and its working principle is provided. In Sec.~\ref{sec:characterization} and Sec.~\ref{sec:Dust charac},  various plasma and dusty plasma diagnostic techniques with main findings are presented, respectively. Some of the identified upcoming experiments with their novelty are highlighted in Sec.~\ref{sec:Future}. A concluding remark is provided in Sec.~\ref{sec:Concl}.\par
\section{Experimental Set-up and procedure}\label{sec:setup}
\begin{figure}[htp]
\centering{\includegraphics[scale=0.29]{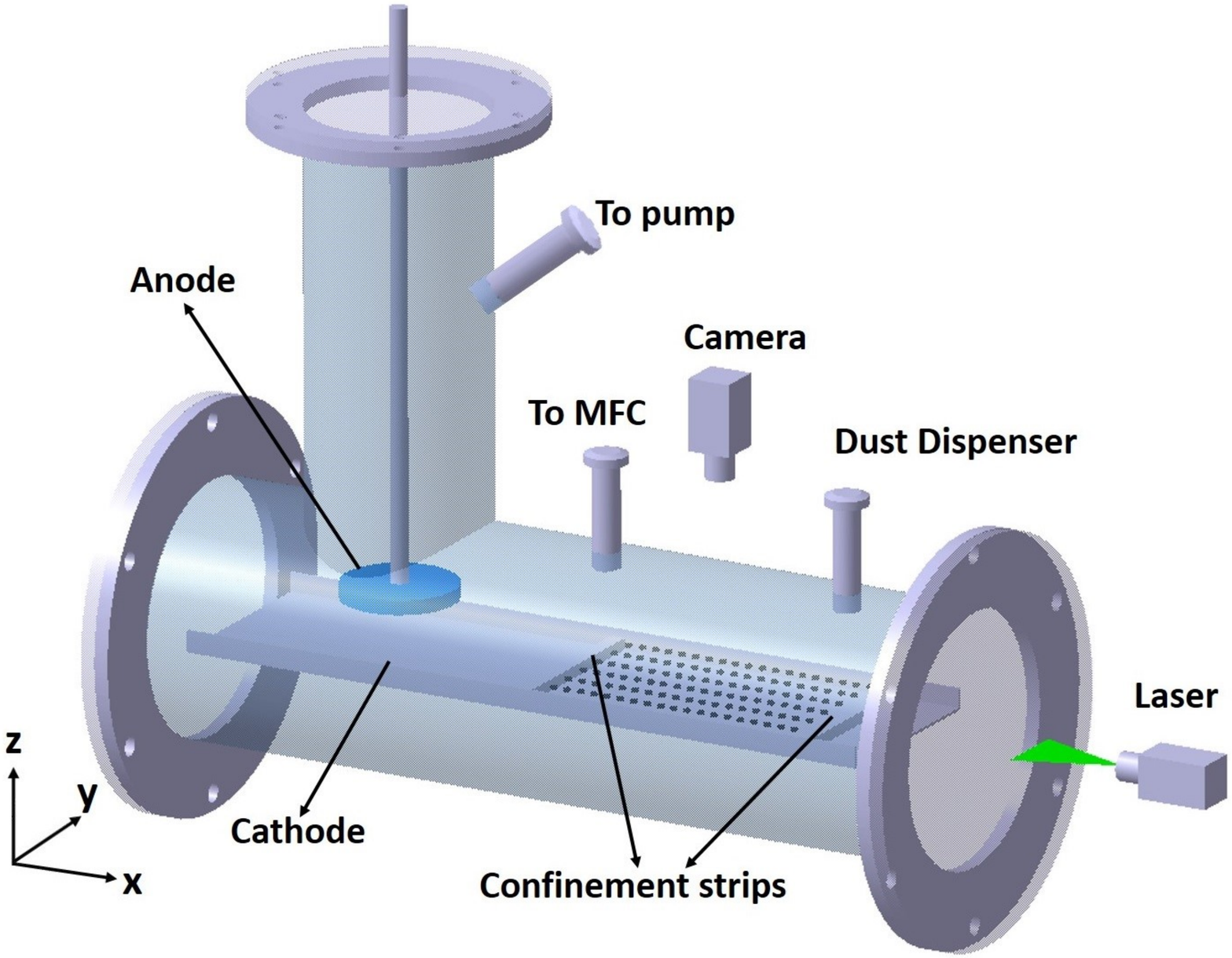}}
\caption{\label{fig:fig1}~Schematic diagram of the DPEx-II device.}
\end{figure}

The schematic of the experimental set-up is shown in Fig.~\ref{fig:fig1}. This tabletop dusty plasma experimental device consists of an L-shaped pyrex glass tube with a number of axial and radial ports to connect to different accessories. The length of the horizontal channel is 60~cm whereas the inner diameter (ID) is 15~cm. The length and ID of the vertical channel are 58~cm and 10~cm, respectively. The circular disc-shaped SS anode having a diameter of 5 cm is inserted from the top most port of the vertical channel. The rectangular tray shaped cathode made of SS is kept horizontally at the mid-plane of the horizontal channel at a distance of 10~cm from the anode. The dimensions of the cathode are 40~cm$\times$15~cm$\times$2~mm which makes its area forty times larger than the area of the anode. In a symmetric electrode discharge configuration with equal electrode areas, the flux of electrons at the anode becomes equal to the flux of ions at the cathode. In contrast, for an asymmetric electrode configuration with larger cathode area (in case of present experiments), the ion flux at the cathode becomes considerably lower than the electron flux at the anode. As a result, the ions enter into the cathode sheath with lower velocity and impart their momentum to the dust particles. Due to this reason, the ion heating effect on the dust particles reduces significantly by a fraction of anode to cathode areas in the case of asymmetric electrode configuration \cite{Lieber}. Therefore, in this work, a highly asymmetric electrode arrangement has been adopted as a primary design requisite of the DC dusty plasma experimental set-up. To further reduce the ion heating effect the cathode geometry is additionally modified by folding the edges of the cathode to a height of about $2$ cm in the $z$ direction throughout its length. This unique arrangement helps to reduce the ion streaming at the center and increases towards the folded edges of the cathode. This fact is verified by examining the prolonged ion exposed surface features of the cathode. Near to the folded edges, dark burnt areas are appeared due to ion bombardment, whereas at the center such burning spots are not observed \cite{Hari1}.\par
To begin with, the chamber is evacuated by a rotary pump to attain its base pressure of 0.001~mbar and the pressure inside the chamber is monitored by two Pirani gauges. One of them is connected at the top port of the horizontal channel and the other is connected at the vertical channel. To achieve the working pressure ($p$) of $\approx$ 0.10~mbar, Argon gas is introduced into the chamber through a mass flow controller (MFC) by setting up the appropriate flow rate. Argon plasma inside the chamber is produced by applying a DC voltage ($V_d$) of 325~V between the anode and cathode through a current limiting resistor of 2 k$\Omega$. The single and double Langmuir probe and emissive probes are inserted from one of the topmost port of the horizontal channel for plasma characterization. Mono-dispersed Melamine Formaldehyde (MF) dust particles having diameter of 10.66 $\mu m$ and mass density of 1.5 g/cm$^3$ are introduced by a dust dispenser. As soon as dust particles enter into the plasma, the higher flux of electrons than ions from the plasma to the dust particles makes them negatively charged. These negatively charged dust particles keep accelerating towards the cathode surface due to gravity. When the dust particle enters into the cathode sheath region, it experiences a counteracting electrostatic force. The balance of these two forces makes the dust particles settle in one plane parallel to the cathode. A couple of SS rectangular strips separated by a distance of 14~cm is used to provide the axial confinement (along x-direction) to the dust particles whereas the curved edges of the cathode provide radial confinement (along y-direction). The dust particles are illuminated by a thin laser line source and the Mie scattered light from the dust particles is captured by a CCD camera placed perpendicular to the dust cloud. The high-resolution images/videos from the camera are stored in the computer to study the dynamics of the dust particles \cite{DPEX}.
\section{Plasma Characterization}\label{sec:characterization}
 After the creation of direct current glow discharge argon plasma in the device, rigorous plasma characterization is made using various electrostatic probes e.g. single and double Langmuir probes and an emissive probe. Due to the larger dimension, the plasma parameters (electron temperature, plasma density, plasma potential \textit{etc.}) certainly differ in this device compare to DPEx device even if the discharge conditions are kept similar. Therefore, it is essential to characterize the DC glow discharge plasmas and Coulomb crystals in this newly build DPEx-II device over a range of discharge conditions for upcoming experiments. The single and double Langmuir probes are mainly employed to measure the electron density and temperature, whereas the emissive probe is used to measure the plasma potential. The probes are scanned along the axial (along $x$ from the left edge of the cathode) and radial (along $z$ from the surface of the cathode) directions to obtain axial and radial profiles of the plasma parameters. The details of the characterization techniques along with the results are presented in the following sub-sections.
\subsection{Single Langmuir Probe Measurements}
Single Langmuir probe is one of the well known and simplest plasma diagnostic techniques widely used for plasma diagnostics. It can be used to find plasma parameters such as plasma and floating potentials, electron temperature, plasma density, electron energy distribution function (EEDF), \textit{etc.} \cite{lang1,lang2}. In this study, a cylindrical Langmuir probe of length of 1~cm and diameter of 1~mm is used to determine the electron density and temperature over a range of discharge conditions.
A ramp voltage is applied to this single Langmuir probe above the floating potential. Firstly, the electron temperature is obtained from the inverse of the slope of the transition region of the $\textit{I-V}$ curve. The plasma density is obtained from the ion saturation current with the help of estimated electron temperature. It is found that the electron temperature and the plasma density varies from 1.5 -- 4.0~eV and $1\times10^{15}-5\times10^{15}$~m$^{-3}$ for the pressure range of 0.10~mbar to 0.14~mbar and the voltage range of 275~V -- 375~V, respectively. \par
\begin{figure}[ht]
\centering{\includegraphics[scale=0.54]{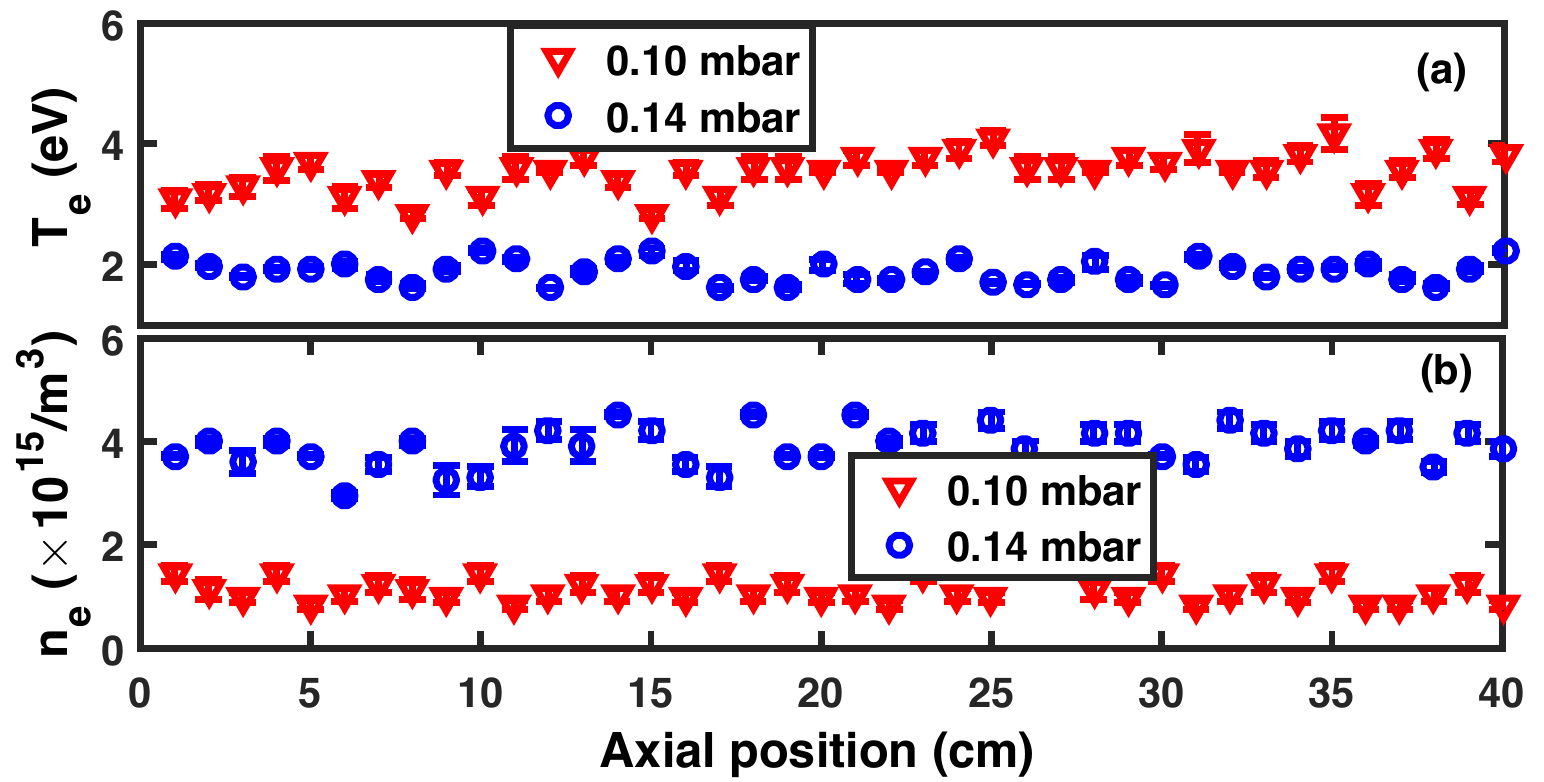}}
\caption{\label{fig:fig2}~Axial profiles of (a) electron temperature and (b) plasma density obtained from single Langmuir probe measurements for the discharge voltage of 375~V and discharge pressures of 0.100~mbar and 0.140~mbar.}
\end{figure}
To obtain the axial profiles of plasma density and electron temperature, the single Langmuir probe is axially scanned (along x) at a height of 5 cm from the cathode surface over a range of discharge conditions. Fig.~\ref{fig:fig2}(a) shows the axial profile of electron temperature whereas Fig.~\ref{fig:fig2}(b) shows the axial plasma density profile. As can be seen from Fig.~\ref{fig:fig2}(a) the value of electron temperature remains almost constant along the axis of the chamber for a given discharge voltage of 375~V and background pressure of 0.100~mbar (open triangle) and 0.140~mbar (open circle). With increased collisionality with the neutrals at higher background pressure, the electron temperature becomes lower. The plasma density remains approximately constant for the same discharge condition. Due to the higher degree of ionization, the plasma density shows a higher value at higher background pressure for a given discharge voltage of 375~V.
These measurements essentially establish that the plasma in our device remains almost uniform along the axis of the chamber for a given discharge condition.
\subsection{Double Langmuir Probe Measurements}
We have also used a double Langmuir probe to measure the plasma parameters for characterizing the DC glow discharge plasma. The double probe is superior to the single Langmuir probe as it does not draw much current from the plasma and as a result causes less local plasma perturbations \cite{double1}. To validate the measurements of the single Langmuir probe, a series of experiments have been carried out to estimate the electron temperature and the plasma density using the double Langmuir probe. In this present set of experiments, two tungsten wires of length 1.0~cm and diameter of 1.0~mm separated by a distance of 1.0~cm are used as a double Langmuir probe. For the range of plasma parameters estimated from the single Langmuir probe, the electron Debye length comes out to be in the range of 0.15 -- 0.35~mm, which is much smaller than the probe separation. For the $\textit{I-V}$ characteristics, one probe is biased from -15~V to 15~V  with respect to another probe, and the current is measured across a resistor of 1 k$\Omega$ connected in series with the probe.
\begin{figure}[ht]
\centering{\includegraphics[scale=0.64]{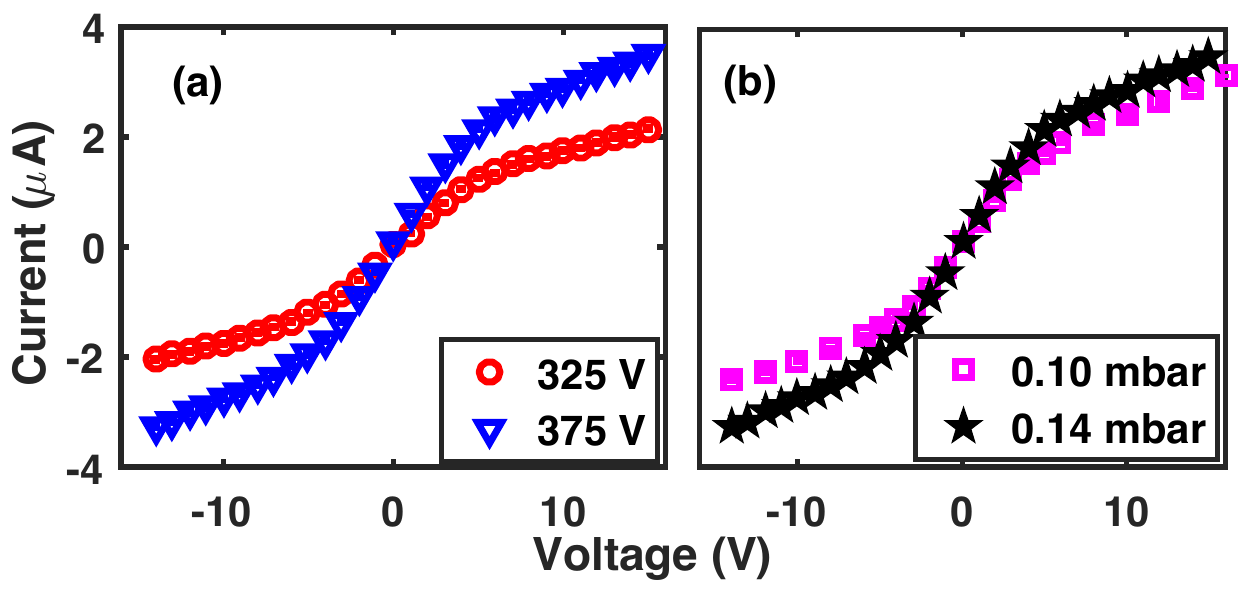}}
\caption{\label{fig:fig3} $\textit{I-V}$ characteristics of double Langmuir probe obtained for (a) $V_d=$~325~V and $V_d=$~375~V for a fixed pressure of 0.14~mbar and (b) for the pressures of 0.10~mbar and 0.14~mbar at fixed $V_d=$~375~V at a radial distance of $1$~cm from the cathode surface.}
\end{figure}
Figure~\ref{fig:fig3}(a) shows the typical $\textit{I-V}$ characteristics curve of the double Langmuir probe for $V_d=325$~V and 375~V at $p=0.1$~mbar whereas Fig.~\ref{fig:fig3}(b) shows the same for two different gas pressures of $p=0.10$~mbar and 0.14~mbar at a discharge voltage $V_d=375$~V. One can see from the figure that unlike a single Langmuir probe, the double Langmuir probe $\textit{I-V}$ characteristic is symmetric about the voltage axis. The plasma density is estimated from the ion saturation current whereas the electron temperature is estimated from the logarithmic slope of the transition region similar to the single Langmuir probe. It is evident from Fig.~\ref{fig:fig3} that higher the value of $V_d$ and $p$ higher is the ion saturation current and lower the slope of the transition region of the $\textit{I-V}$ characteristics. \par
\begin{figure}[ht]
\centering{\includegraphics[scale=0.64]{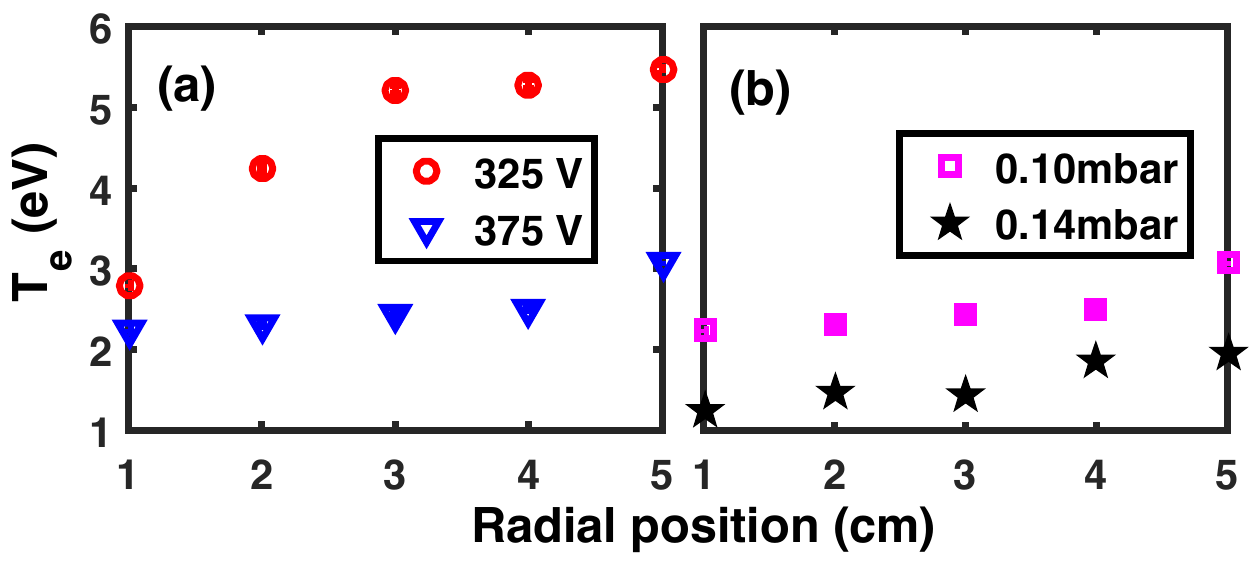}}
\caption{\label{fig:fig4} (a) Estimated radial electron temperature profiles for two different $V_d$ of 325~V and 375~V at a fixed $p$ of 0.14~mbar and (b) radial electron temperature profile for two different pressures of 0.10~mbar and 0.14~mbar for a constant value of $V_d=375$~V using double Langmuir probe.}
\end{figure}
For the range of discharge conditions, the electron temperature is estimated from the obtained $\textit{I-V}$ characteristics. The estimated values of electron temperature along the radial direction of the chamber are shown in Fig.~\ref{fig:fig4}. Figure~\ref{fig:fig4}(a) shows the trend of radial $T_e$ obtained for two different $V_d$ of 375~V and 325~V at the $p$ of 0.1~mbar. Figure~\ref{fig:fig4}(b) shows the radial electron temperature profile for two different values of $p$, namely, 0.10~mbar and 0.14~mbar for  $V_d$ of 375~V. It can be noted from Fig.~\ref{fig:fig4} that with the increase of radial distance from the cathode surface, the electron temperature increases. It is also worth mentioning that at a particular radial location, the electron temperature is higher for lower discharge voltage and neutral gas pressure. The reduced number of collisions between electrons and background neutrals makes the electron temperature higher at lower $p$. In order to maintain the constant plasma pressure the electron temperature increases with the decrease of $V_d$ for a given $p$.\par
\begin{figure}[ht]
\centering{\includegraphics[scale=0.64]{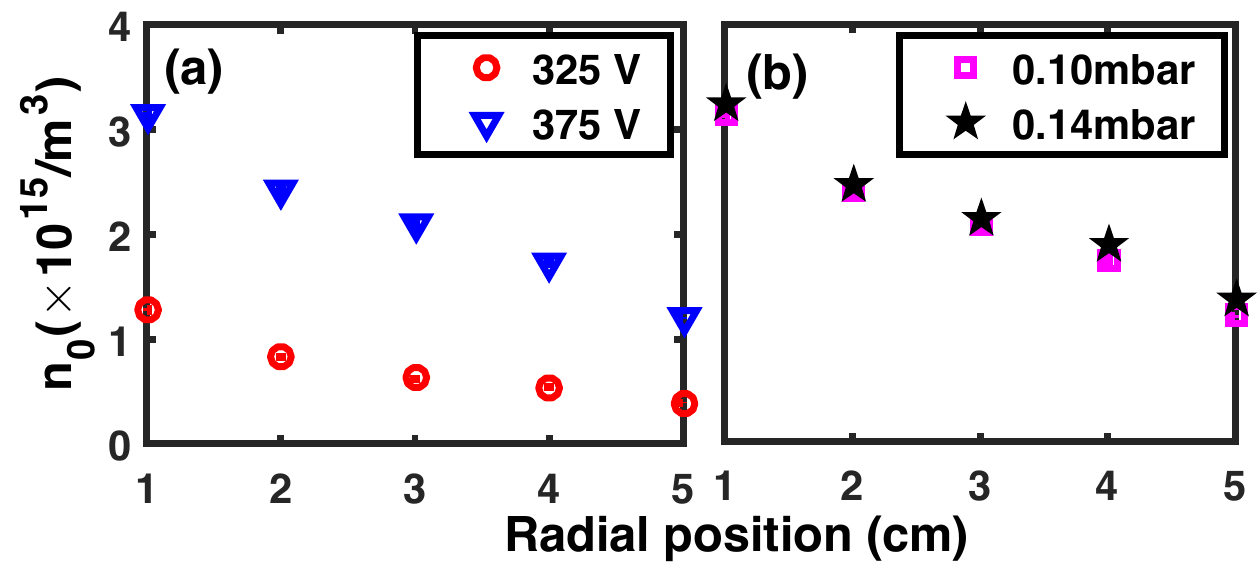}}
\caption{\label{fig:fig5} Estimated radial plasma density profiles for two different $V_d$ of 325~V and 375~V at a fixed $p$ of 0.14~mbar and (b) radial density profiles for two different $p=0.10$~mbar and $p=0.14$~mbar for a fixed $V_d$ of 375~V using double Langmuir probe.}
\end{figure}
An attempt is also made to estimate the plasma density from the ion saturation current over the same range of discharge conditions for which the value of $T_e$ is measured. Figure~\ref{fig:fig5} shows the radial profile of plasma densities. It is clear from Fig.~\ref{fig:fig5}, that the plasma density is higher for higher $p$ and $V_d$ at a particular radial location. The plasma density is maximum at the cathode sheath edge and it decreases monotonically when the probes are scanned away from the cathode due to the diffusive nature of the plasma. The radial increase of electron temperature (as shown in Fig.~\ref{fig:fig4}) is due to the  decrease of plasma density (as shown in Fig.~\ref{fig:fig5}) to maintain the constant plasma pressure in the radial direction. It can be inferred that the plasma parameters estimated using a double Langmuir probe show a good agreement with the plasma parameters measured with the help of a single Langmuir probe at similar discharge conditions. \par
\subsection{Emissive Probe Measurements}
Accurate estimation of plasma potential in the radial direction is essential for determining the position of the levitation of dust particles, whereas the axial plasma potential profile determines the spatial distribution of dust particles in the axial direction. Precise measurements of plasma potential using an emissive probe inside the sheath make it one of the important plasma diagnostic techniques \cite{emissive1,emissive2}.  As a consequence, it is also employed to find the radial and axial plasma potential profiles for the present set of experiments.  A hairpin shaped tungsten wire having a diameter of 0.125~mm and length of 1~cm with a ceramic holder is used to make the emissive probe. An emissive probe can be used to measure the plasma potential using three different methods such as floating-point, separation point, and inflection point methods \cite{emissive1,emissive2}. For a particular discharge condition, the plasma potential is estimated using all these three different methods and found to lie close to each other within the error limit. Hence, we have adopted the easiest technique of floating-point method to measure the plasma potential using the emissive probe.\par
\begin{figure}[ht]
\centering{\includegraphics[scale=0.64]{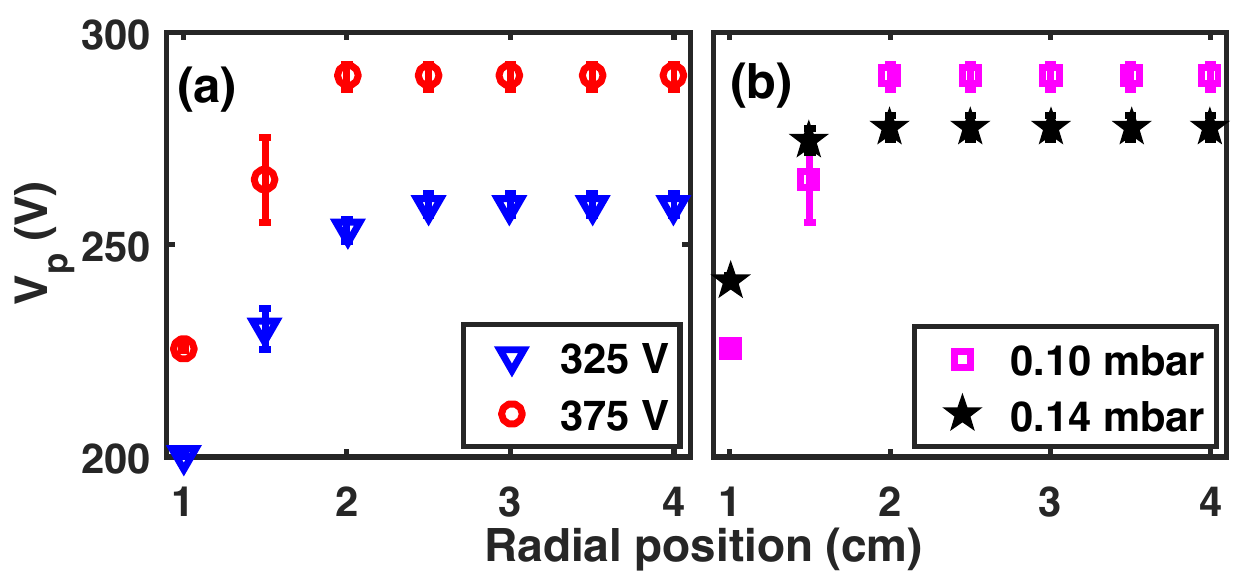}}
\caption{\label{fig:fig6} (a) Radial profile of plasma potentials obtained from emissive probe for the $p$ of 0.1~mbar at two different $V_d$ of 325 V and 375 V. (b) Radial profile of plasma potentials for two different $p$ of 0.10~mbar and 0.14~mbar at the $V_d$ of 375 V.}
\end{figure}
Firstly, the emissive probe is used to measure the plasma potential radially by varying the discharge voltage at 325~V and 375~V for a background $p$ of 0.1~mbar. Figure~\ref{fig:fig6}(a) shows such radial variation of plasma potential for these two discharge voltages. It is worth mentioning in Fig.~\ref{fig:fig6}(a), the position of the probe at 1.0~cm indicates the cathode sheath edge. It is clear from the figure that near the cathode, the plasma potential is less, and as we go away from the cathode along the radial direction, the plasma potential increases. This drop-in plasma potential near the cathode surface is due to the  presence of cathode sheath. The increase of discharge voltage at a particular radial location results in an increase of plasma potential due to the presence of more energetic electrons. Figure~\ref{fig:fig6}(b) shows the radial plasma potential profile for a $V_d$ of 325~V at two different $p$ of 0.10 and 0.14~mbar. It is found that the higher the value of $p$ lower the plasma potential for the given value of $V_d$. The frequent collisions at higher pressure makes the electrons lose their energy resulting in a decrease of plasma potential. The drop of plasma potential near the cathode sheath region can be used to estimate the electrostatic force that causes the dust particles to levitate. \par
\begin{figure}[ht]
\centering{\includegraphics[scale=0.53]{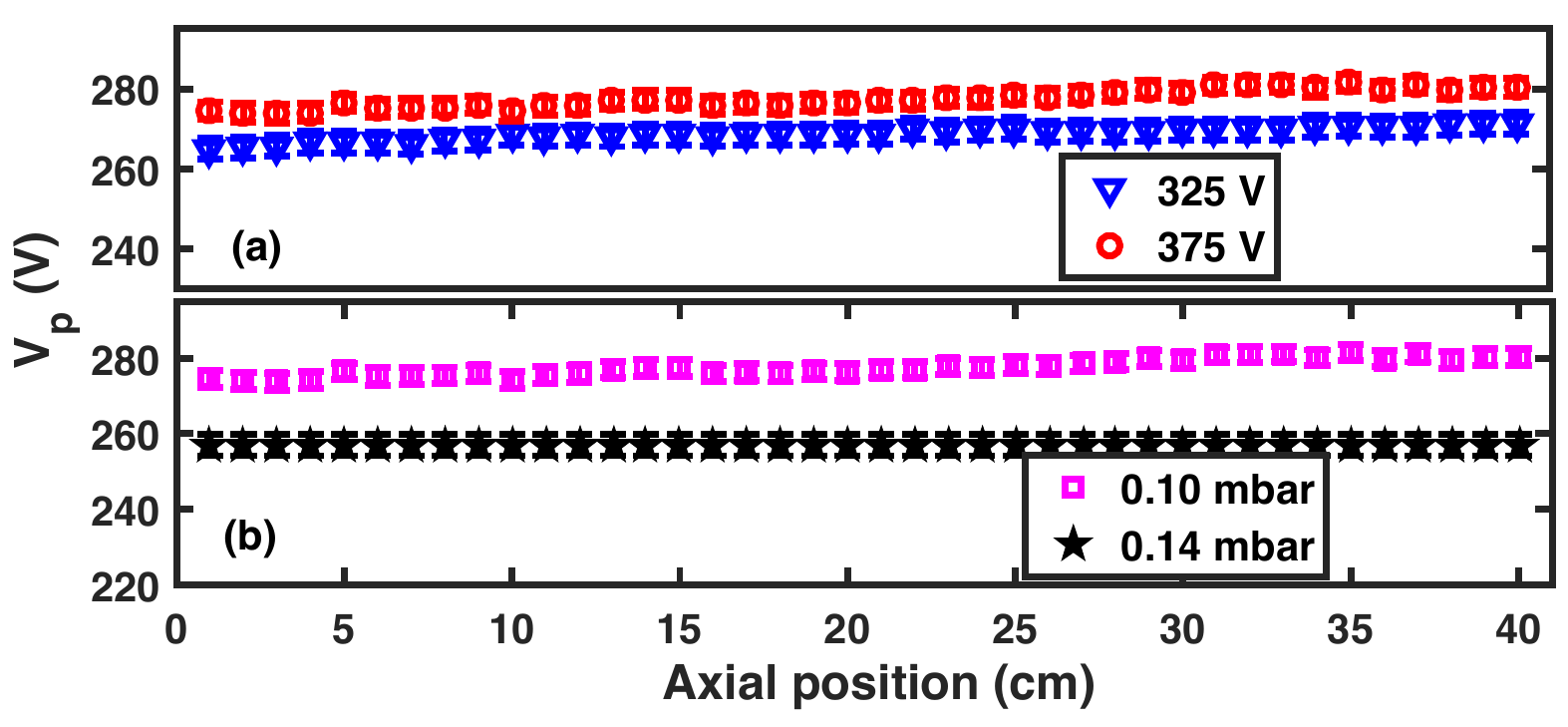}}
\caption{\label{fig:fig7} (a) Axial profile of plasma potential obtained from the emissive probe for the $p$ of 0.1~mbar and for two different $V_d$ of 325 V and 375 V. (b) Axial profile of plasma potential for the $V_d$ of 325 V and two different $p$ of 0.10~mbar and 0.14~mbar.}
\end{figure}
An attempt is also made to measure the axial plasma potential profile using the same floating-point method. Figure~\ref{fig:fig7}(a) shows the axial plasma potential profile obtained for two different $V_d$ of 325 and 375~V at a $p$ of 0.10~mbar. Similarly, Fig.~\ref{fig:fig7}(b) shows the axial plasma potential profile obtained for the $V_d$ of 375 V at two different $p$ of 0.10 and~0.14~mbar. It is found that the axial plasma potential remains uniform for a given discharge condition. The variation of plasma potential with the discharge parameters at a given axial location shows a similar trend as that of the radial profiles of plasma potential. Thus, the uniformity of plasma potential along the axial direction allows us to form a long homogenous dust crystal in the present device.\par
\section{Characterization of Dust Crystal}\label{sec:Dust charac}
After characterizing the plasma thoroughly, Melamine Formaldehyde (MF) dust particles having a diameter of 10.66~$\mu m$ and mass density of 1.5 $g/cm^3$ are introduced by a dust dispenser installed at one of the ports of the horizontal channel (see Fig.~\ref{fig:fig1}). As the dust particles enter into the plasma, they absorb more electrons than ions (due to the higher mobility of electrons) and get negatively charged. These particles get fully charged when they fall through the plasma due to gravitational force by traveling a distance of a few cm as the charging time is few orders lower in magnitude than the time of flight. When they reach near  the cathode sheath region, they feel a repulsive electrostatic force due to the cathode sheath electric field of $\approx1.2\times10^4$~V/m as discussed in Sec.~\ref{sec:characterization}. These dust particles settle at one X-Y plane at a particular height (Z) where counteracting gravitational force balances the electrostatic force. Using the force balance condition, one can find the dust charge ($Q_d$) for a given size of the dust particles and the electric field. For our experimental conditions, the estimated value of $Q_d$ comes out to be 4840$e$ for the particle size of 10.66 $\mu$m and the electric field strength of $1.2\times10^4$~V/m, which is in fair agreement with the values reported in past \cite{dustcharge1,dustcharge2}.
\begin{figure}[ht]
\centering{\includegraphics[scale=0.40]{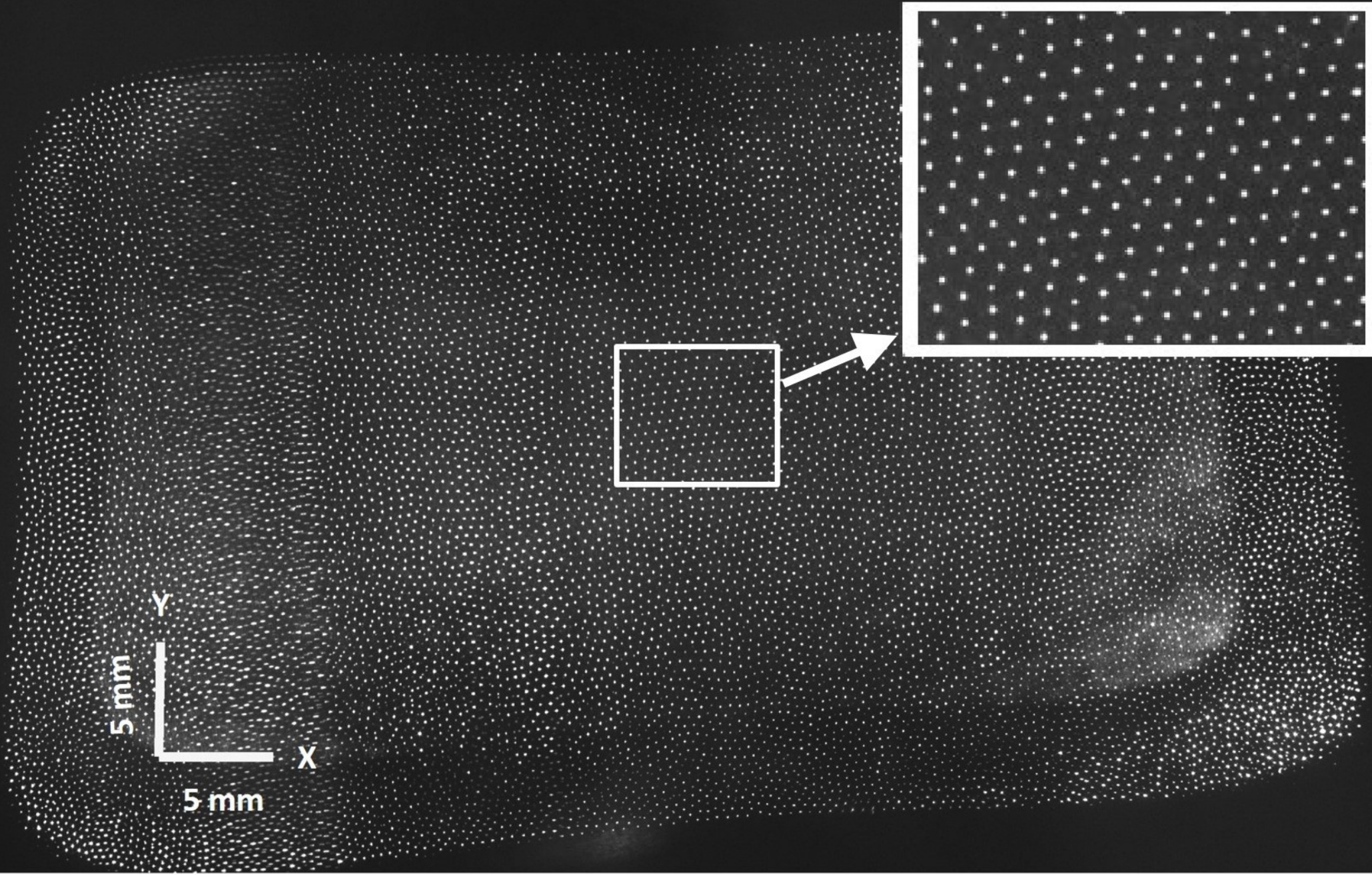}}
\caption{\label{fig:fig8} A large dust crystal of length 8~cm and width of 4~cm obtained at a discharge voltage of 310~V and a pressure of 0.040~mbar.}
\end{figure}
To overcome the repulsion between the like-charged particles, a couple of stainless steel (SS) strips are placed on the cathode separated by an approximate distance of 10 cm for the horizontal confinement. As mentioned in Sec.~\ref{sec:setup}, a line laser light is used to illuminate the particles in the X-Y plane and the scattered light is captured by a CCD camera. Figure~\ref{fig:fig8} shows a typical image of a dust crystal having the length of 8~cm and width of 4~cm at a $V_d=310$~V and a $p=0.040$~mbar. An enlarged view of the central region (the region enclosed by the square) of the crystalline structure is shown in the inset of Fig.~\ref{fig:fig8}. The crystalline structure can be made bigger by placing the confining strips at a larger distance. It is worth mentioning that the uniform nature of DC background plasma (inferred from the results of different plasma diagnostic techniques) supports the formation of a large size (up to approximately the dimension of cathode) homogeneous dust crystal in our device. The structural and dynamical properties of this large size dusty plasma crystal are investigated through a host of characteristic parameter estimations such as radial pair correlation function, Voronoi diagram, Delaunay Triangulation, and the dust temperature, which will be discussed in the following sub-sections.
\subsection{Pair Correlation Function}
For a system of \textit{N} particles, the radial pair correlation function $g(r)$ gives the information on how the particles are distributed in space from the position of a reference particle in that system. The radial pair correlation function is used in our experiment to estimate the inter-particle distance, density and the range of order \cite{Hari3, Hari1, DPEX} for a given crystal structure.
\begin{figure}[ht]
\centering{\includegraphics[scale=0.25]{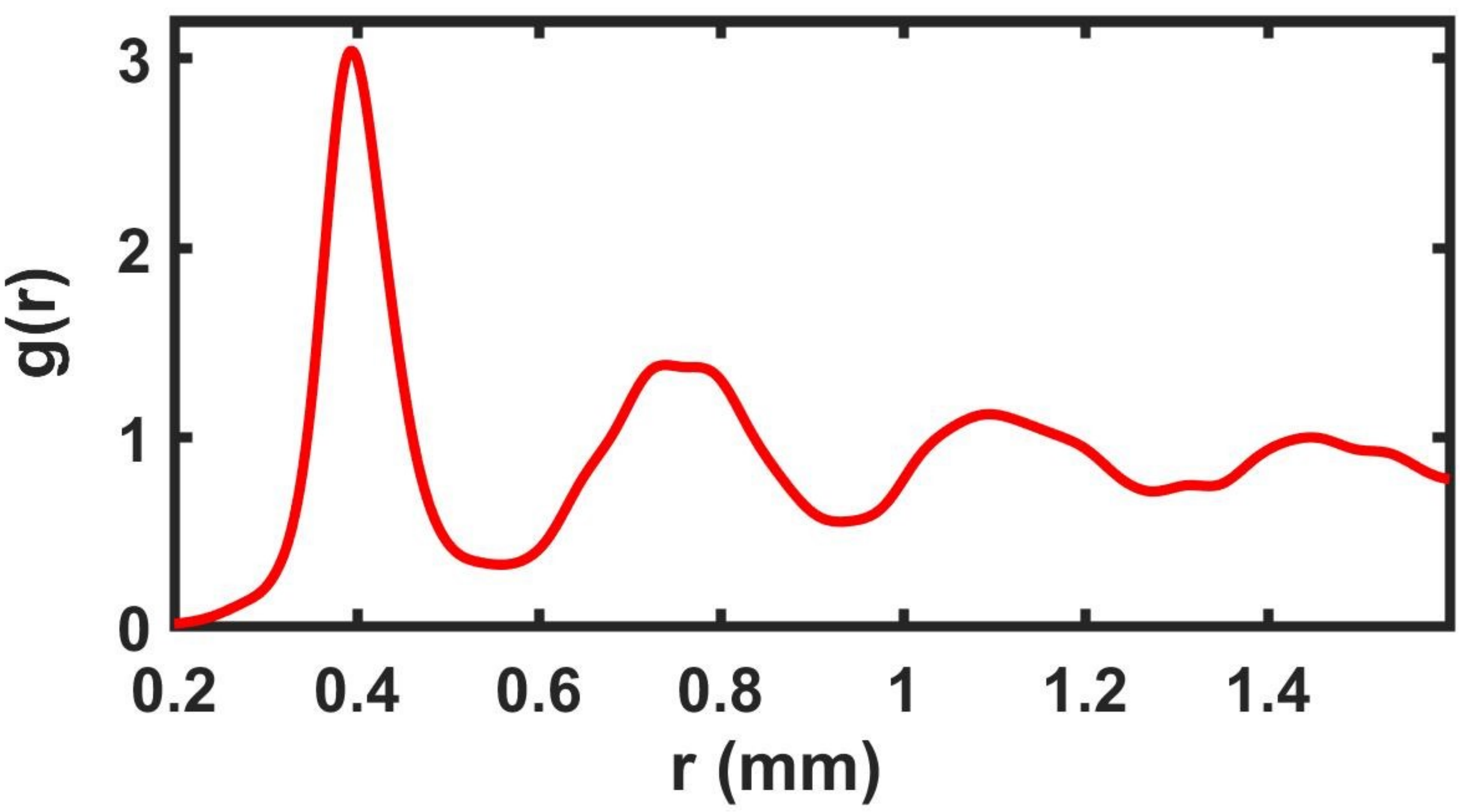}}
\caption{\label{fig:fig9} Radial pair correlation function of the dust crystal formed at $p=0.040$~mbar and $V_d=310$~V. }
\end{figure}
Figure~\ref{fig:fig9} shows the typical pair correlation function extracted from the coordinates of the particles encircled in Fig.~\ref{fig:fig8}. The position of the first peak as shown in Fig.~\ref{fig:fig9} gives the information about the average inter-particle distance by which all the particles are distributed in the dust crystal. For this specific discharge condition, the inter-particle distance and the dust density comes out to be $\sim 435$ $\mu$m and $1.6\times10^6$/m$^3$, respectively. The multiple peaks in Fig.~\ref{fig:fig9} gives also an indication of long-range order in the crystal. Further, we extend the pair correlation function analysis over the complete crystal to find the spatial distribution of inter-particle distance and dust density by dividing it into small segments separated by a distance of 0.5~$cm$. Figure~\ref{fig:fig10} shows the estimated inter-particle distance and density along the length of the dust crystal. We found that the value of inter-particle distance and density remain almost constant throughout the length of the dust crystal.\par
\begin{figure}[ht]
\centering{\includegraphics[scale=0.58]{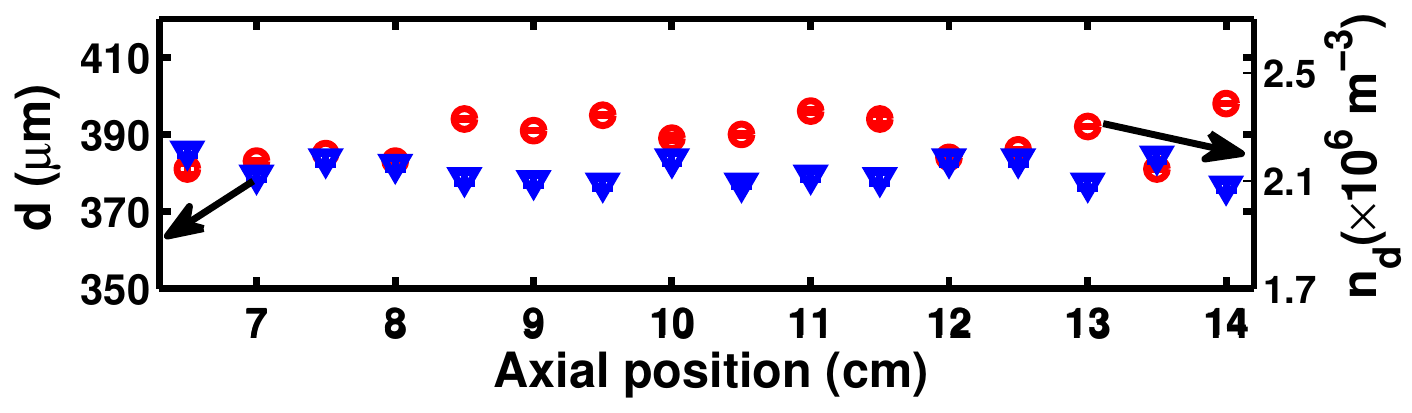}}
\caption{\label{fig:fig10}Axial profiles of inter-particle distance and density over the dust crystal obtained at the neutral gas pressure of 0.035~mbar and discharge voltage of 310~V.}
\end{figure}
\begin{figure}[ht]
\centering{\includegraphics[scale=0.27]{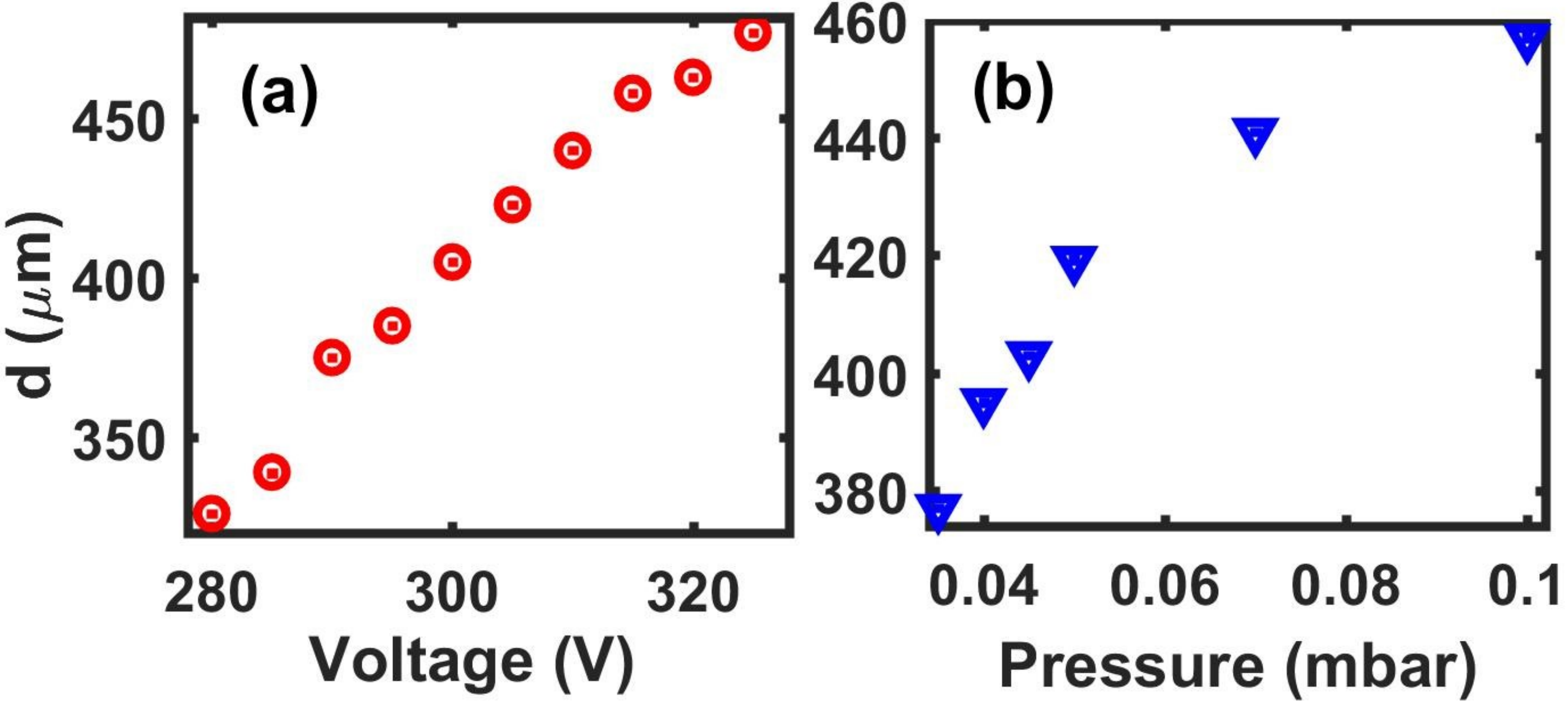}}
\caption{\label{fig:fig11}(a) Inter-particle distance as function of voltage for $p=0.04$~mbar and (b) Inter-particle distance as function of $p$ for a fixed $V_d=310$~V.}
\end{figure}
\begin{figure}[ht]
\centering{\includegraphics[scale=0.60]{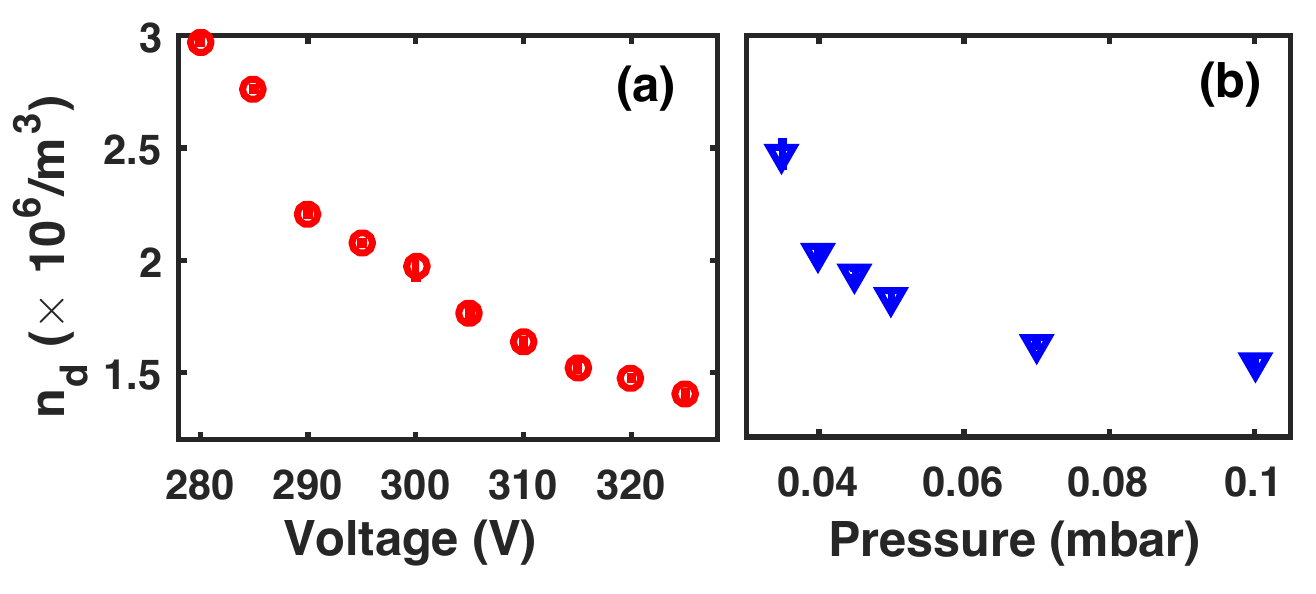}}
\caption{\label{fig:fig12}(a) Variation of dust density as a function of discharge voltage for $p=0.04$~mbar and (b) Variation of dust density as a function of $p$ for a fixed $V_d=310$~V.}
\end{figure}

The dusty plasma crystal is then characterized over a range of discharge conditions by measuring the inter-particle distance and the particle density. Figure~\ref{fig:fig11}(a) shows the variation of inter-particle distance as a function of discharge voltage at a fixed $p$ of 0.04~mbar, whereas Fig.~\ref{fig:fig11}(b) shows the same as a function of $p$ for a given discharge voltage of 310~V. The value of inter-particle distance was found to  increase with an increase of both $V_d$ and $P$. From the information of inter-particle distance, we attempted to estimate the dust particle density over the same range of discharge conditions.
The estimated value of number density as a function discharge voltage is shown in Fig.~\ref{fig:fig12}(a) at a fixed $p$ of 0.04~mbar and Fig.~\ref{fig:fig12}(b) shows the number density of dust particle as a function of $p$ for a fixed discharge voltage of 310 V.
It is clear from the above analysis that for the increase of both $V_d$ and $p$, the value of inter-particle distance increases and number density decreases. With the increase of both $V_d$ and $p$, the plasma density increases, resulting in a decrease of the electron Debye length. It causes a decrease in the sheath thickness around the confining strips, which ultimately allows the particles to expand due to the inter-particle repulsive force among them. As a result, the inter-particle distance increases and the dust density decreases with the increase of $V_d$ and $p$.
\subsection{Voronoi Diagram and Delaunay Triangulation}
 The Voronoi diagram \cite{Delaunay1, Delaunay2} gives a convenient way to quantify the amount of order and hence disorder in a 2-D crystalline structure. In dusty plasmas, it is often used to find the degree of crystallization of a dust crystal \cite{Hari3}.
\begin{figure}[ht]
\centering{\includegraphics[scale=0.6]{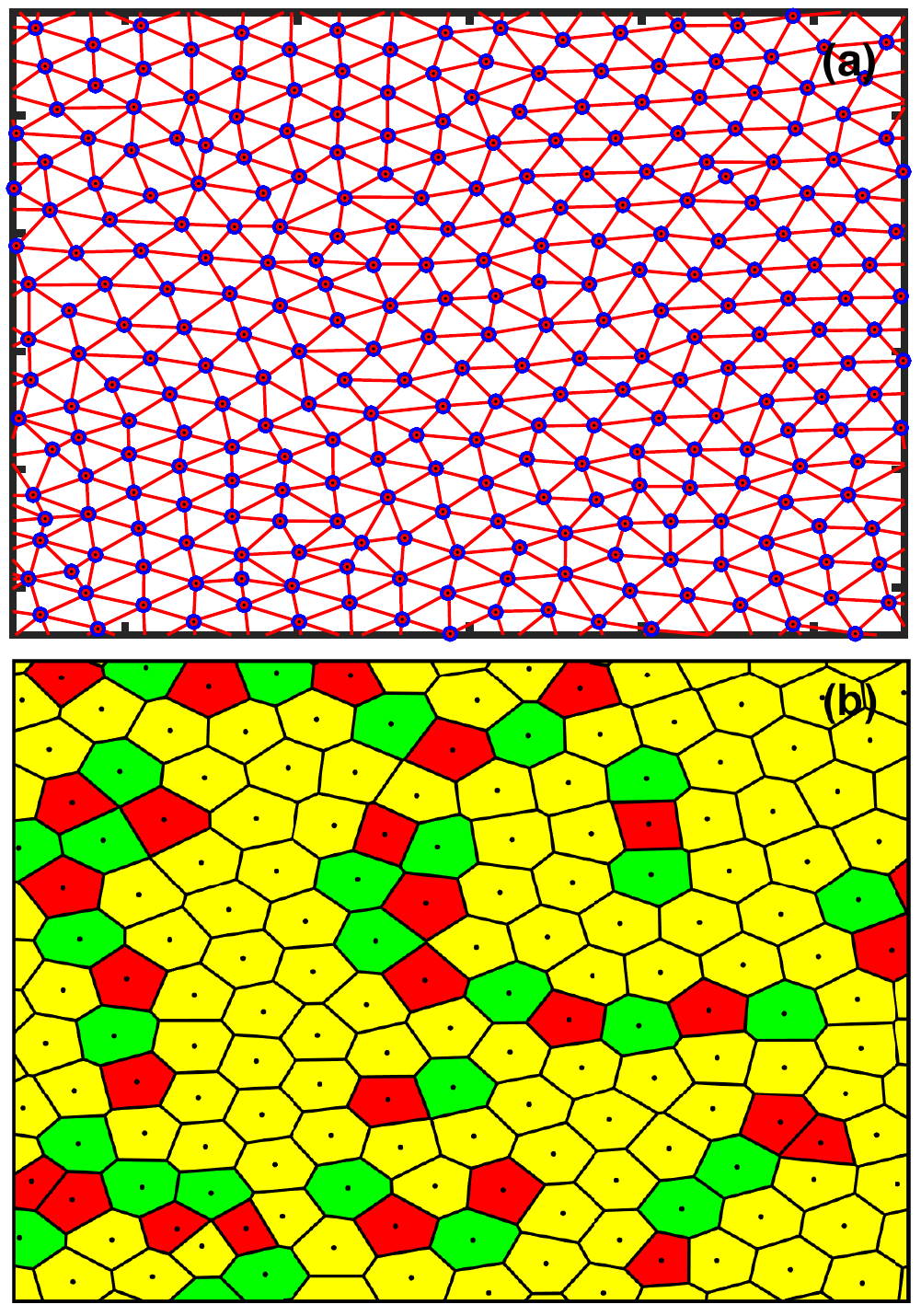}}
\caption{\label{fig:fig13}(a) Delaunay triangulation and (b) Voronoi diagram obtained for the crystal structure shown in the encircled region of Fig.~\ref{fig:fig8}.}
\end{figure}
 Implementation of the Voronoi diagram can be done easily through Delaunay triangulation which forms a network of triangles by connecting all the nearest neighbors to a given point in the plane of a crystal \cite{Delaunay1, Delaunay2}. For the construction of the Delaunay triangles and Voronoi diagram, the encircled area of Fig.~\ref{fig:fig8} is used. Figure~\ref{fig:fig13}(a) shows the Delaunay triangulation whereas Fig.~\ref{fig:fig13}(b) depicts the corresponding Voronoi diagram of the same crystal. The non-parallel lines in the Delaunay triangulation as shown in Fig.~\ref{fig:fig13}(a) essentially indicate the region of defects in the crystalline structure. In the Voronoi diagram as shown in Fig.~\ref{fig:fig13}(b), the ordered hexagonal region is filled by yellow color while the defect regions are filled by green and red colors. All the particles in the yellow shaded region have six nearest neighbors and hence form hexagonal cells whereas the non-hexagonal cells are accounted as defects.  It is to be noted that the hexagonal cells are the most favorable structure to form a purely 2-D crystalline structure in the case of dusty plasma also \cite{packing}. The structural order parameter (P) of a crystal structure is defined as the ratio of the total number of the hexagon ($N_H$) to the total number of polygons ($N_P$) in the crystal structure as \(P=\frac{N_H}{N_P}\times100\) \cite{Hari1, Hari3}. The value of $P$ obtained for the crystal structure shown in Fig.~\ref{fig:fig8} is 88.7\%.
\subsection{Measurement of Dust Temperature}
 The temperature of a physical system is one of the important thermodynamical quantities to estimate various dynamical parameters associated with it. To characterize the dust crystal in the present set of experiments, an attempt is made to estimate the dust temperature from the information of the coordinates of the individual particles in the crystal over a sufficient span of time. The average random thermal velocity of the dust particles in a crystalline structure is quantified as dust temperature \cite{Hari1, Hari3}. The dust particles attain a thermal equilibrium by undergoing frequent collisions with background ions and neutrals and follow a velocity distribution function, which is assumed as a Gaussian velocity distribution function \cite{Hari1}.
 \begin{figure}[ht]
\centering{\includegraphics[scale=0.27]{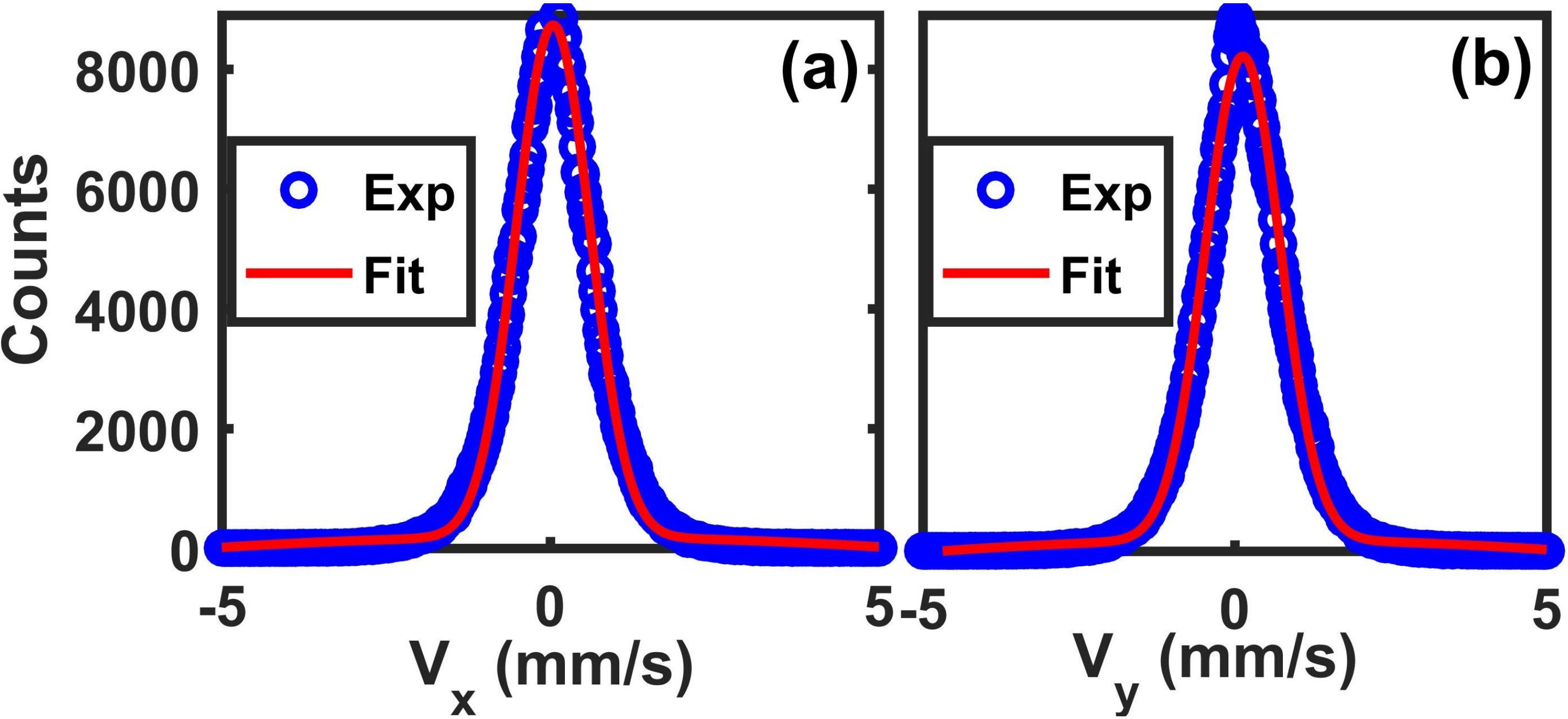}}
\caption{\label{fig:fig14}Velocity distribution of dust particle along (a) x-direction and (b) y-direction.}
\end{figure}
 The standard deviation of this Gaussian distribution function gives an estimate of the dust temperature. The open circles in Fig.~\ref{fig:fig14}(a) and \ref{fig:fig14}(b) show the experimentally obtained velocity distribution of the dust particles along $x$ and $y$-directions for the discharge voltage of 310~V and pressure of 0.04~mbar for the encircled region of Fig.~\ref{fig:fig8}. The solid lines represent the best fit of the Gaussian distribution functions. The dust temperature is approximately estimated from these distribution functions and comes out to be about 2.0~eV for the same discharge condition. To examine the uniformity of the dust temperature in the crystal, we extended the temperature measurement over the length of the dust crystal by dividing it into small segments separated by a distance of 0.5~$cm$. Figure~\ref{fig:fig15} shows the variation of dust temperature along the length of the dust crystal for $V_d=310$~V and $p=0.035$~mbar. The location \lq0\rq (not shown in Fig.~\ref{fig:fig15}) corresponds to the edge of the cathode. It is clear from the figure that the dust temperature of the crystal remains almost constant along the axial extent.\par
 \begin{figure}[ht]
\centering{\includegraphics[scale=0.58]{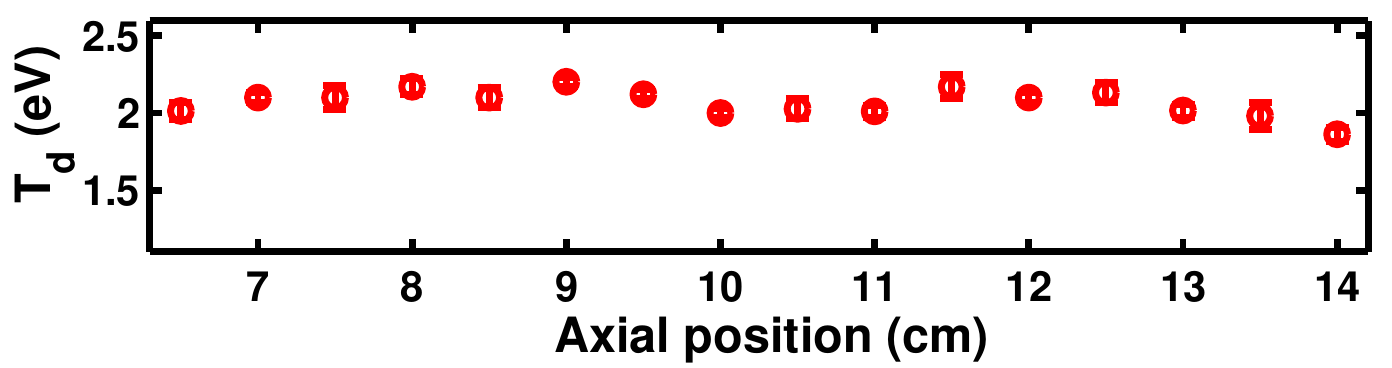}}
\caption{\label{fig:fig15}Axial profile of dust particle temperature for the dust crystal obtained at the pressure of 0.035~mbar and discharge voltage of 310~V.}
\end{figure}
\begin{figure}[ht]
\centering{\includegraphics[scale=0.60]{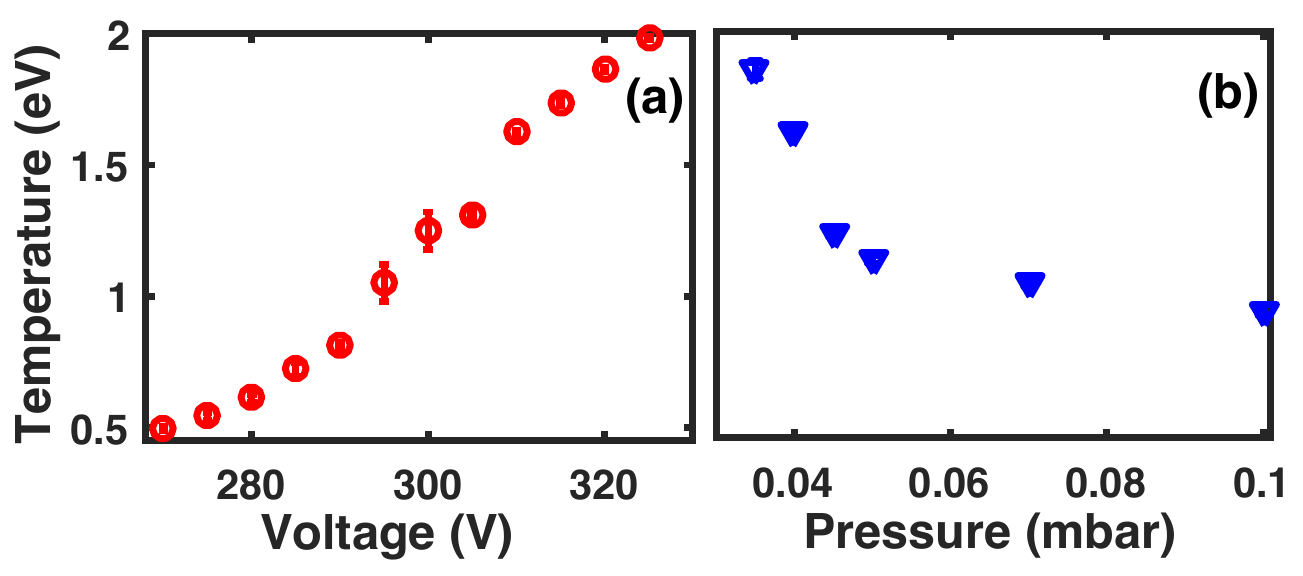}}
\caption{\label{fig:fig16}Variation of dust temperature with (a) $V_d$ at a fixed pressure of 0.040~mabr and (b) neutral gas pressure for a fixed $V_d=310$~V.}
\end{figure}

Further, we carried out the same measurement over a range of discharge conditions at a given axial location of x=10~cm. The variation of dust temperature as a function of $V_d$ for a fixed discharge pressure of 0.04~mbar is shown in Fig.~\ref{fig:fig16}(a) whereas the dust temperature as a function of $p$ at a given $V_d$ of 310~V is shown in Fig.~\ref{fig:fig16}(b). As can be seen from Fig.~\ref{fig:fig16} that with an increase in discharge voltage the dust temperature increases but with the increase of $p$, the crystal temperature decreases.
Further, increasing of $V_d$ at fixed $p=0.04$~mbar leads to a decrease in Debye length and hence the decrease in confinement sheath thickness. Hence, the dust particles expand more due to their mutual repulsive force which causes the dust temperature to increase at higher $V_d$. The frequent collisions of dust particles with the neutral gas molecules increase with the increase of $p$ at a fixed $V_d=310$~V leads to decrease in the dust temperature. \par
\section{Future Studies}\label{sec:Future}
This table top versatile DC dusty plasma set-up is built to perform experiments in strongly coupled dusty plasma over a larger volume. It is evident from the above characterization that the present dusty plasma experimental set-up allows us to form a larger size, stable, and uniform dust crystal. Another interesting advantage of the present system is that it supports confining dust crystals using a variety of confining potential structures. A series of experiments have already been carried out in this device and some of the initial findings are discussed in the following subsections.  However, the details of the following experiments are not the main focus of the present study and have been listed to demonstrate the versatility and experimental potential of the DPEx-II device.
\subsection{Different Crystal Confinement Structure}
As our present DC dusty plasma experimental set-up has a larger cathode surface area, it allows the formation of a dust crystal using a variety of confinement geometry over a range of sizes from a few $cm^2$ to a few hundred $cm^2$ of area.
\begin{figure}[ht]
\centering{\includegraphics[scale=0.27]{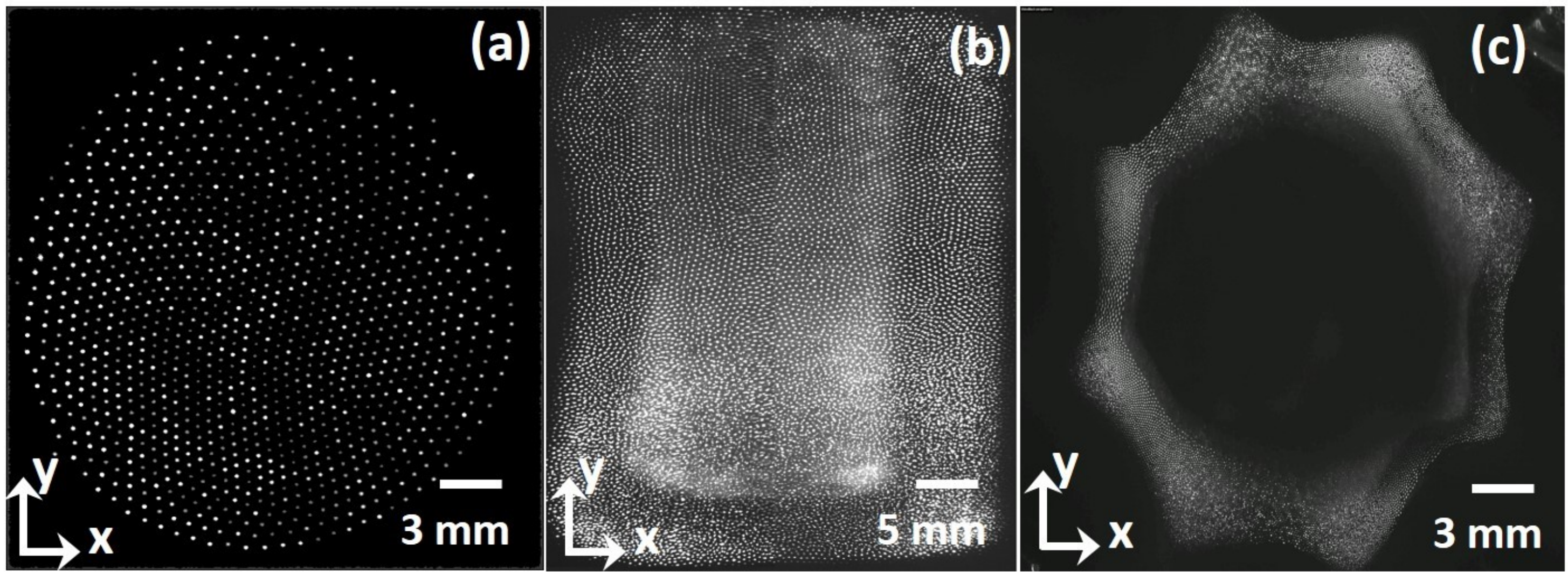}}
\caption{\label{fig:fig17} Dust crystal formation using (a) circular, (b) rectangular and (c) saw-teeth potential structures respectively over a range of $V_d$ starting from 300~V - 450~V and $p$ from 0.035~mbar to 0.14~mbar.}
\end{figure}
Fig.~\ref{fig:fig17}(a), (b), and (c) shows dust crystals formed in this device using circular, rectangular, and saw-teeth potential structures respectively over a range of $V_d$ starting from 300 V - 450 V and $p$ from 0.035~mbar to 0.14~mbar. We used circular confinement for the study of particle level dynamics whereas the rectangular confinement structures is used for the investigation of wave dynamics in a dusty plasma medium. One can also use saw-teeth potential structure to study the non-equilibrium statistical features of dusty plasma. Thus, the facility of handling multiple confinement structures helps to perform various experiments and the associated unexplored physics in this device.
\subsection{Void Formation}
A void is a dust-free region in dusty plasma that has a characteristic sharp boundary maintained by two counteracting forces: electrostatic and the ion drag force \cite{void3,void1,void2,RF_Void,Pro_ind_void,Polyakov_2017,Vrancken_2005,van_de_Wetering_2015,Mitic_2013}. The ion drag force acts on the dust particles in outward direction and builds a dust free region. On the other hand, the electrostatic force acts on the negatively charged dust particle in inward direction (opposite to the ion drag force), which hinders the extend of the formation of void structure. The radius of the void structure depends on the magnitude of these two forces. Void and its formation mechanism have been extensively studied in RF discharge \cite{void3,RF_Void,Pro_ind_void,Vrancken_2005,van_de_Wetering_2015,Mitic_2013}. There are only a few studies performed purely in DC discharges where the void is formed in the standing strata created in the positive column \cite{void1,void2,Polyakov_2017}.
\begin{figure}[ht]
\centering{\includegraphics[scale=0.27]{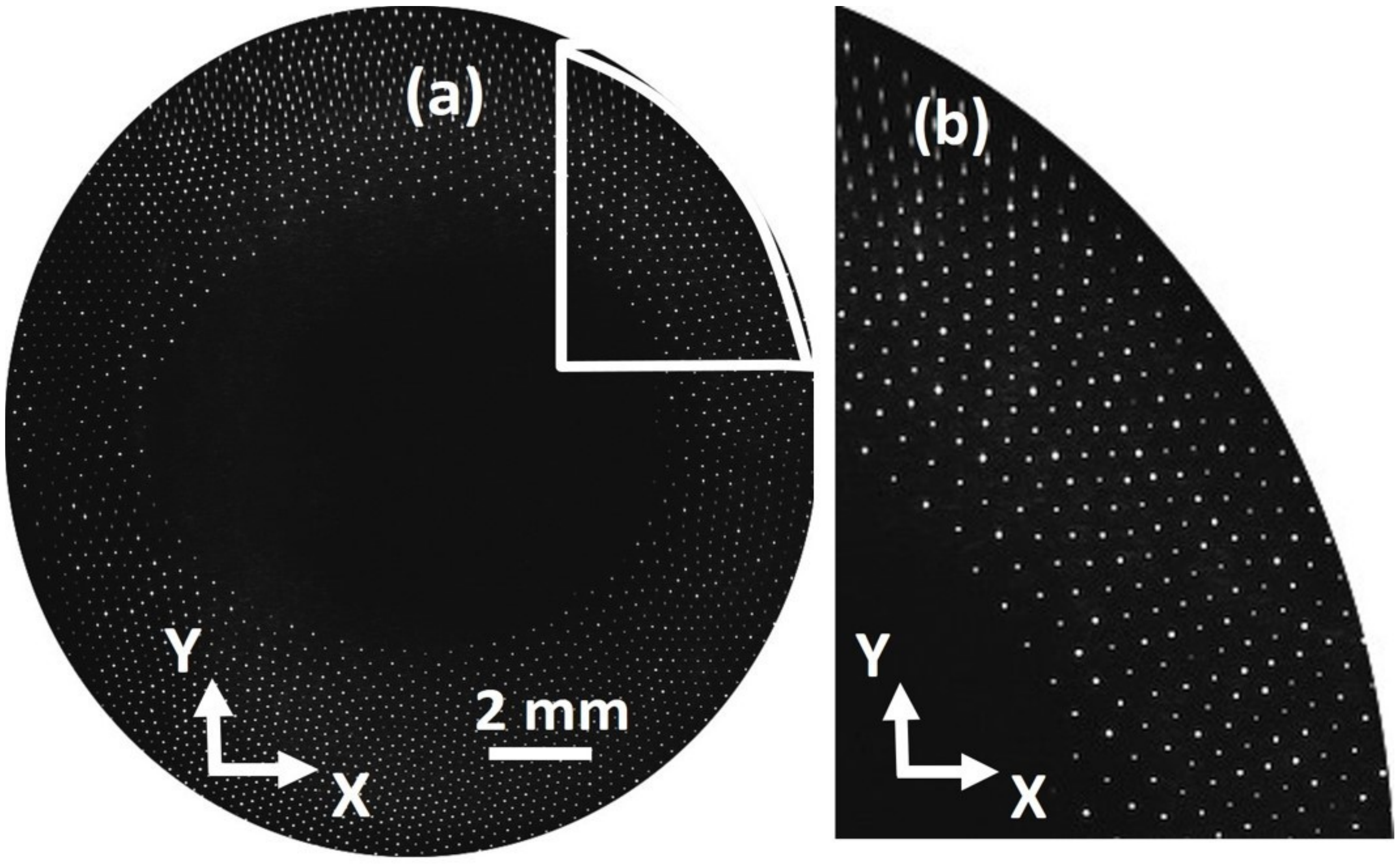}}
\caption{\label{fig:fig18} (a) Void structure obtained in DPEx-II device for the discharge voltage of 350~V and pressure of 0.12~mbar. (b) Zoomed view of the void structure at one of the corners.}
\end{figure}
The formation of a void in such geometries is restricted over a narrow range of discharge conditions. Figure~\ref{fig:fig18}(a) shows the formation of the void structure obtained in circular confinement geometry in the present set of experiments at a discharge voltage of 350~V and background pressure of 0.12~mbar.  It is worth mentioning that the dimension of void and its radius increases with the increase of $V_d$ and $p$. The void structures in this asymmetric electrode configuration are highly repeatable, stable, and symmetric over a wide range of discharge conditions. Other than circular geometry, the voids are also seen in rectangular and saw-teeth like confinement geometries. Figure~\ref{fig:fig18}(b) depicts the zoomed view of one of the corners of the void structure. The zoomed view clearly shows the periodic arrangement of dust crystals with a sharp boundary between the crystal and dust-free region.
\subsection{Void Rotation}
During the last couple of decades, an extensive investigation on the rotation of dust particles has been carried out in different experimental conditions \cite{Thomas, Mangilal, Manjit}. The dust particles exhibit rotation in the background of plasma due to crossed electric and magnetic fields \cite{Thomas}. In absence of an externally applied magnetic field, the dust particles are also found to rotate both in RF and DC plasmas \cite{Mangilal, Manjit}. The emergence of the directed motion of a colloidal system is often been explained in the context of non-equilibrium statistical mechanics as ratchets. Experimental realization of ratchet in various physical systems such as colloidal particle confined in a 1-D optical trap in water medium \cite{Fey_Rat_Exp}, rotational ratchet in granular gases \cite{Rotat_ratchet}, ratchet in doped semiconductor structure \cite{Semi_ratchet} are extensively studied. However, in past, the rotation of dust crystal or realization of dusty plasma ratchet in DC glow discharge plasma has not been investigated. In the demonstration of a ratchet mechanism, the system needs to possess a spatially asymmetric potential profile in the engaged state between ratchet and pawl, and it has to be in contact with one or two reservoirs that supply random force for its Brownian motion. As the dust particles in a dusty plasma are electrically charged, hence all the above-stated criteria can be achieved straightforwardly similar to other systems.\par
\begin{figure}[ht]
\centering{\includegraphics[scale=0.27]{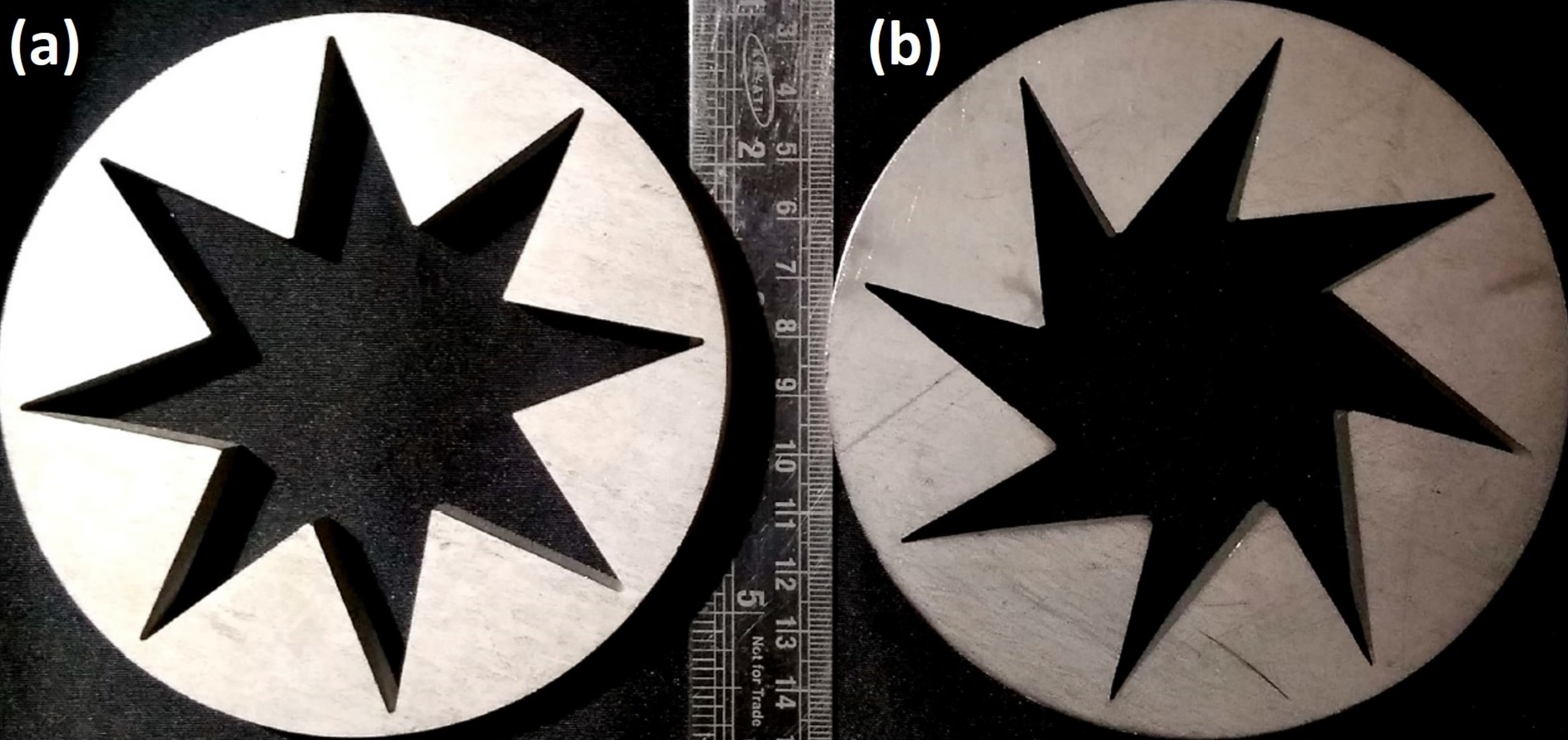}}
\caption{\label{fig:fig19} Photograph of spatially periodic (a) symmetric and (b) asymmetric gears having 8 number of saw-teeth used for void rotation.}
\end{figure}
\begin{figure}[ht]
\centering{\includegraphics[scale=0.25]{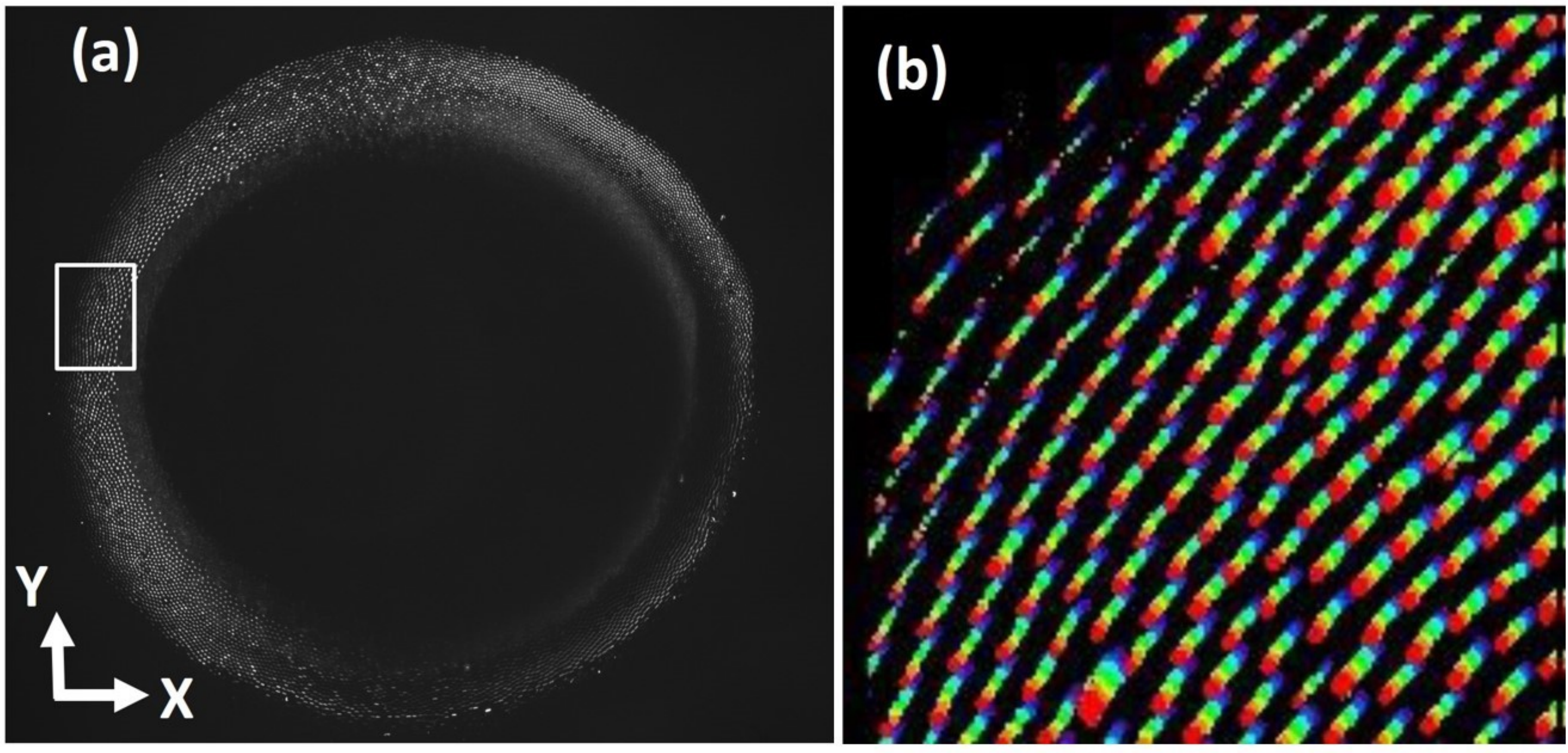}}
\caption{\label{fig:fig20}(a) Typical image of void rotation and (b) overlapped positions of particles over a span of time of 200~msec while rotating in counter-clockwise direction for $V_d=400$~V and $p$ of 0.1~mbar. The blue color indicates the initial positions of the dust particles, whereas the red color depicts their final positions. }
\end{figure}

In the present set of experiments, the saw-teeth shaped gears serve the periodic symmetric and asymmetric potentials in the plasma environment as shown in Fig.~\ref{fig:fig19}. The experiments are performed initially with a periodic potential structure by keeping a symmetric saw-teeth gear (see Fig.~\ref{fig:fig19}(a)) having the outer diameter of 14~cm placed on the cathode. In this situation, the dust particles do not display rotation even in the presence of a void of a maximum diameter of 30~mm. It essentially indicates the necessitate of periodic asymmetric potential for the ratching effect. To obtain directed motion, a similar saw-teeth gear of the same dimension with asymmetric structure as shown in  Fig.~\ref{fig:fig19}(b) is used. It provides the necessary spatially periodic asymmetric potential due to the formation of sheath around each saw-tooth in the plasma environment. The ion drag force in the outward direction initiates the formation of a void and acts as a pawl. When the void starts to form, the ion drag force takes the dust particles from the center to the edge of the confining potential well where they feel the influence of spatially periodic asymmetric potential, which results in the rotation of particle as shown in Fig.~\ref{fig:fig20}. The presence and absence of void indicate engaged and disengaged states of the ratchet and pawl, respectively. It is observed, with the increase of discharge voltage and neutral gas pressure, the ion drag force increases and forms the void with a larger dimension \cite{void3}. It yields the dust particles to rotate with higher angular velocity and essentially perform more work.\par
Figure~\ref{fig:fig20}(a) shows such kind of rotation of dust particles for $V_d=400$~V and $p=0.1$~mbar. Fig.~\ref{fig:fig20}(b) shows the overlapped particle positions of one of the corners of the void structure for 200~msec. The blue color indicates the initial positions of the dust particles, whereas the red color depicts their final positions. It is to be noted that the particles located at the outer edge of the void rotate with higher linear velocity than that of the particles located at the inner edge. It essentially signifies that the particles in the void structure show a rigid body rotation in the crystalline state. The direction of rotation of the dust particle solely depends on the direction of potential asymmetricity, which can be switched back by reversing the orientation of saw-teeth gear. For the clockwise periodic asymmetric potential, the dust particles are found to rotate in the clockwise direction whereas for the anticlockwise periodic asymmetric potential, the particles exhibit rotation in the anticlockwise direction.\par
\begin{figure}[ht]
\centering{\includegraphics[scale=0.55]{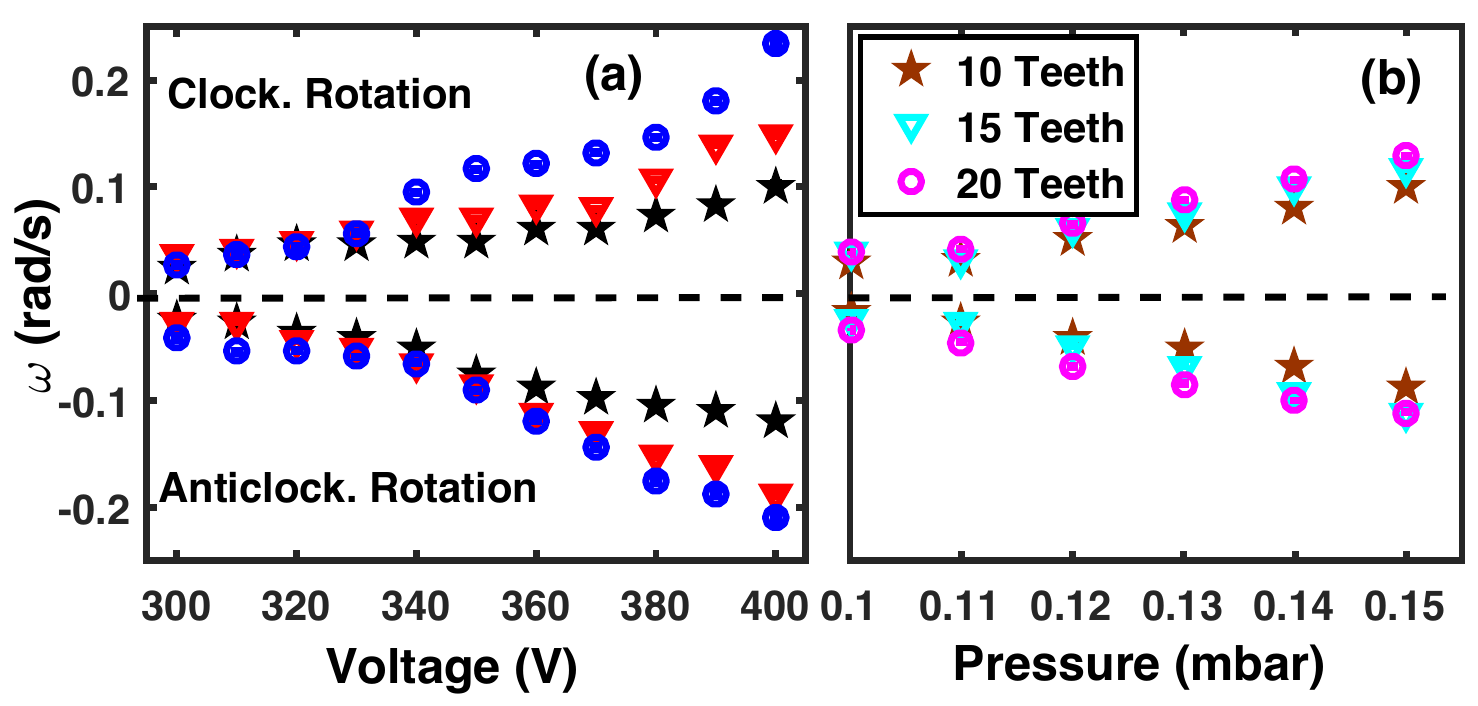}}
\caption{\label{fig:fig21} Variation of average angular velocity of dust particles as a function of (a) discharge voltage at a fixed $p$ of 0.1~mbar and (b) neutral gas pressure for the fixed $V_d$ of 300~V for different confinement structure with teeth number 10 (star), 15 (triangle) and 20 (circle) in clockwise and anticlockwise direction.}
\end{figure}
Figure~\ref{fig:fig21} shows the dependence of the average angular velocity of the dust particles with discharge voltage and background gas pressure for clockwise and anticlockwise rotation with the gears of 10, 15, and 20 teeth. As shown in Fig.~\ref{fig:fig21}(a), the average angular velocity is found to increase from 0~rad/s -- 0.25~rad/s over the range of discharge voltage of 300~V -- 400~V at a fixed value of $p=0.1$~mbar. It is due to the fact that the sheath potential around each tooth becomes stronger with the increase of discharge voltage \cite{Lieber}. Fig.~\ref{fig:fig21}(b) shows the variation of the average angular velocity of the dust particle with the filling gas pressure. The angular velocity increases from 0~rad/s to 0.13~rad/s when the pressure is increased from 0.1~mbar to 0.15~mbar for a fixed $V_d$ of 300~V. At a particular discharge voltage, the sheath around each saw-tooth squeezes with the increase of gas pressure, which results in the increase of the strength of asymmetric potential \cite{Lieber}. The increase of angular velocity with background pressure is not significant as of the discharge voltage. This is because the dust particles undergo frequent collisions with neutrals with the increase of the neutral gas pressure,  which slows down the dust particles. Figure~\ref{fig:fig21} also depicts that the average angular velocity increases for both the cases with the number of saw-teeth structures of the gear at a given discharge voltage or neutral gas pressure. It happens because the strength of the asymmetric potential and ratching frequency increases with the number of saw-teeth structures. Thus the above experimental findings demonstrate that the DC glow discharge dusty plasma in the DPEx-II device serves as an ideal non-equilibrium system to perform rachet experiments.
\subsection{Phase Transitions and Phase Coexistence}
\begin{figure}[ht]
\centering{\includegraphics[scale=0.29]{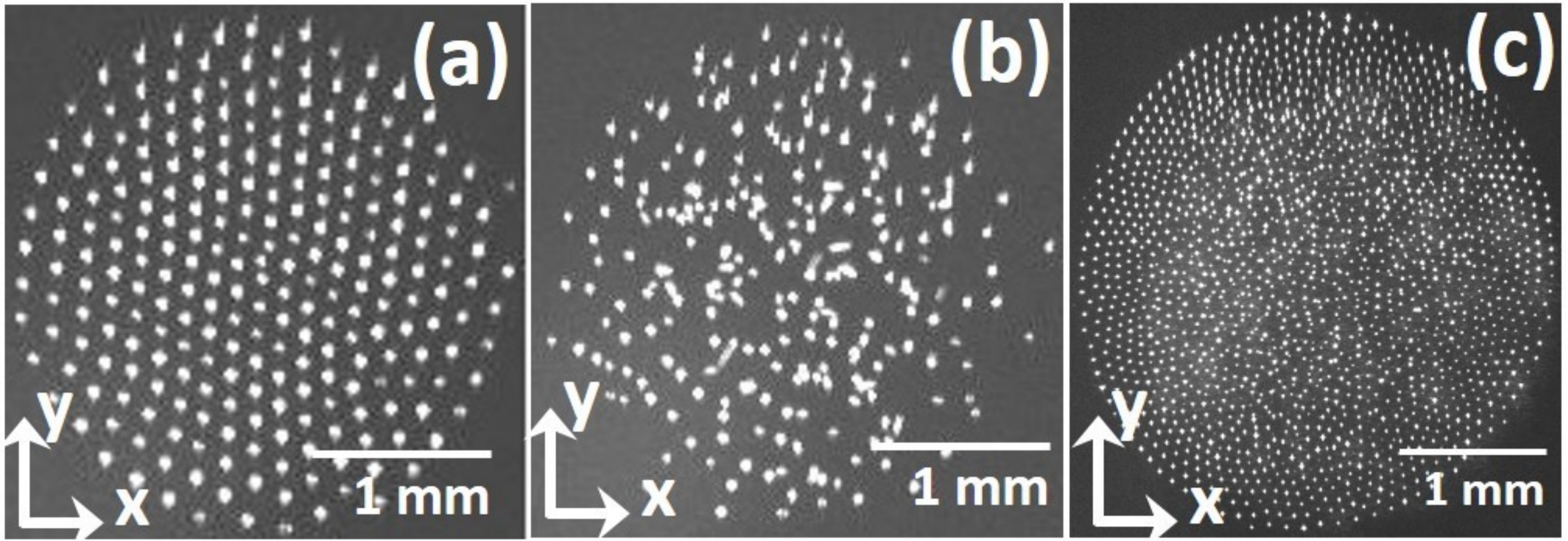}}
\caption{\label{fig:fig22}Photographs of dust particle in (a) crystalline phase (b) liquid phase and (c) in the coexisting crystal and liquid phase.}
\end{figure}

A series of controlled experiments is also been carried out to study the phase transition and phase co-existence in this DC glow discharge device by varying the discharge parameters. Figure~\ref{fig:fig22}(a) shows a crystalline structure that melts completely as shown in Fig.~\ref{fig:fig22}(b) when the $p$ is reduced from 0.077~mbar to 0.066~mbar at a $V_d$ of 480 V.
During the complete phase transition from crystalline state to liquid state, the dust temperature increases significantly from few eV to few tens of eV, whereas the Coulomb coupling parameter decreases by almost an order of magnitude. Strong ion wake induced Schweigart instability is responsible for the melting mechanism in such kind of phase transition \cite{Phasecoex}.\par
In some of the experiments, surprisingly, it is found that the crystalline and liquid states co-exist perpetually for a given discharge condition, which was not evident in past. Fig.~\ref{fig:fig22}(c) shows a typical image of the existence of crystalline phase and the liquid state with the higher number of particles at a $V_d$ of 350 V and $p$ of 0.04~mbar. It is clear from this figure that the liquid phase is surrounded by the crystalline phase. The area of the melted region increases with the decrease in background neutral gas pressure. One of the striking difference between phase coexistence phenomenon in equilibrium and non-equilibrium systems is that there is no force balance mechanism is needed in the later system [54] and, the physical mechanism that triggers this phase separation in non-equilibrium system is not yet understood completely after extensive researches in past [55]. The present system, in which dust particles levitated in the cathode sheath is a well known example of such a non-equilibrium system and can be used as a prototype for the exploration of governing physical mechanism of non-equilibrium phase coexistence phenomenon through controlled experiments.
\subsection{Circular and Spiral Wave Excitation}
Excitations of spiral and circular wave and their propagation have been spotted in varieties of physical system ranging from living cells to astrophysical objects.
\begin{figure}[ht]
\centering{\includegraphics[scale=0.29]{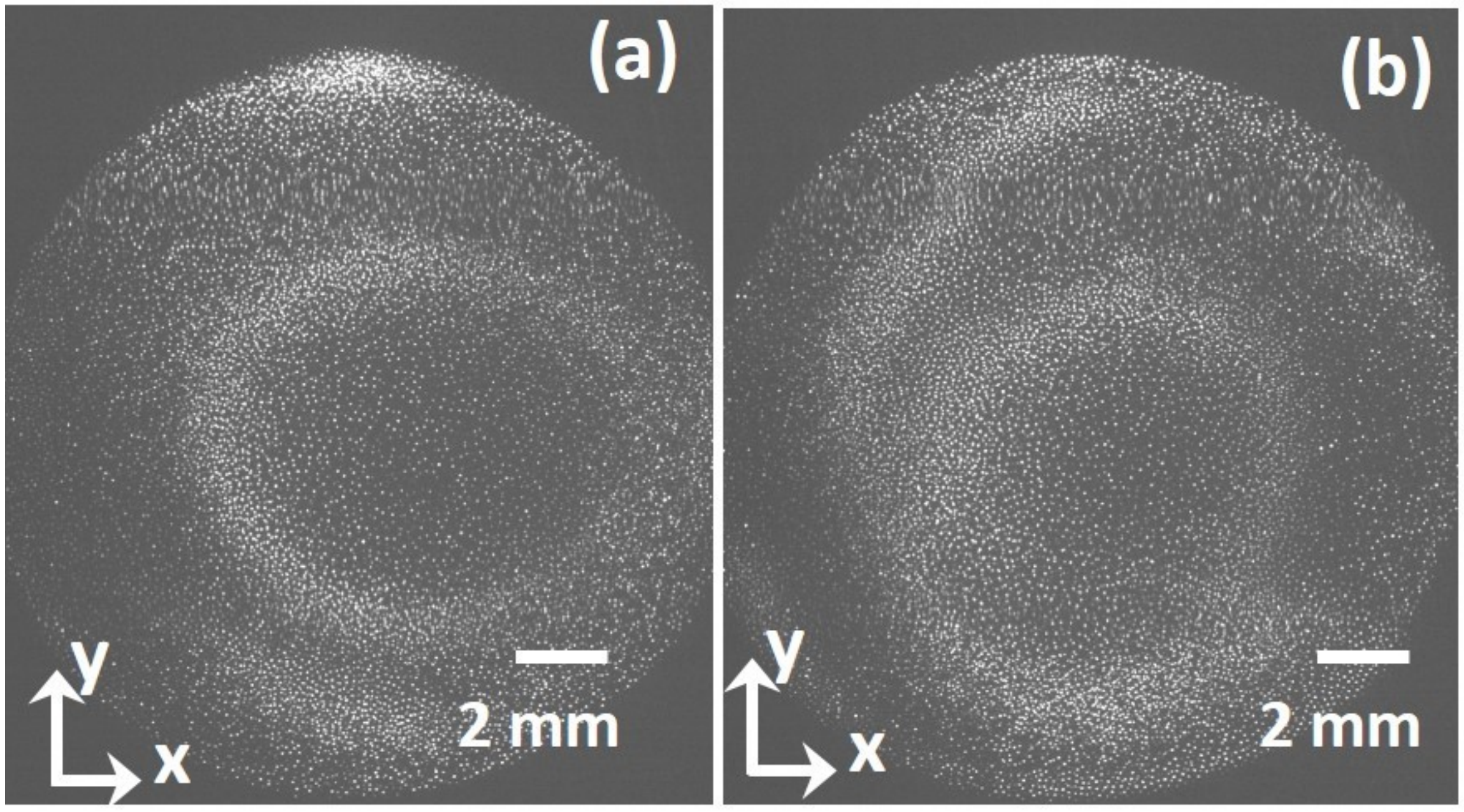}}
\caption{\label{fig:fig23}Photograph of excited (a) circular wave at the $V_d$ of 320 V and at the $p$ of 0.080~mbar (b) spiral wave observed at the $V_d$ of 380~V and the $p$ of 0.080~mbar in a circular potential confinement.}
\end{figure}
The emergence of these waves often been witnessed as an interplay between local inhibitor (prey) and exciter (predator) mechanism in the non-equilibrium systems through computer simulation \cite{spiral_wave,spiral_reac_diff}. The self-organization of excitations in the medium gives rise to the pattern of spiral wave \cite{3D_spiral}. Experimental observation of the excitation of circular and spiral waves have not been reported in dusty plasma so far. An attempt is made to excite the circular and the spiral waves in strongly coupled dusty plasma in this newly built DPEx-II device for the first time. At $V_d$ of 320 V and at $p$ of 0.080~mbar, the spontaneously excited concentric circular waves are observed once when the dust density reaches a threshold density value of $\approx5\times10^{8}$/$m^3$ and found to propagate radially inward. The circular wavefront appears at the inner edge of the confining ring and emanates towards the center as shown in Fig.~\ref{fig:fig23}(a). Upon changing the discharge conditions further, the circular wave transforms into spiral wave as shown in Fig.~\ref{fig:fig23}(b). The details of these experiments will be reported in the upcoming publications.
\section{Conclusions}\label{sec:Concl}
A versatile dusty plasma experimental device is successfully installed at the Institute for Plasma Research (IPR) to perform experiments with large sized dust crystals in a DC glow discharge plasma. The unique feature of this present device is the use of asymmetric electrode configuration which minimizes the heating of dust particles by ion streaming and facilitates the formation of large crystalline structures  in a DC glow discharge plasma. A glow discharge plasmas has already been produced in this newly developed device and characterized thoroughly with the help of various electrostatic probes. The plasma parameters such as plasma density, electron temperature and the plasma potential are found to be in the range of $0.5-4\times 10^{15}$/m$^3$, $1-6$~eV, and $250-280$~V, respectively over the range of discharge voltages of $320-380$~V and of neutral gas pressures of $0.10-0.15$~mbar. The axial profiles of these plasma parameters at a particular radial location remain constant for a given discharge condition. However the plasma density and the electron temperature show a monotonic fall when the measurements are carried out away from the cathode. The plasma potential measured by the emissive probe shows a radial increase near the cathode sheath preceded by a saturation region in the bulk plasma. Mono-dispersive spherical dust particles are introduced in the plasma to create a large sized dust crystal with a maximum area of few hundred cm$^2$, which is later characterized thoroughly by a host of various diagnostic tools. The dust temperature and the dust density lie in the range of 0.5-2.0~eV and $1-3\times 10^{6}$/m$^3$, respectively over a range of discharge conditions and they remain almost constant all over the crystal at a given discharge voltage and pressure. These measurements essentially demonstrate that the large sized dusty plasma crystal produced in this device is homogeneous in nature. A series of initial experiments on crystal rotation, void formation, phase co-existence, circular and the excitations of spiral waves \textit{etc.} have been carried out in this device and their details will be reported in upcoming publications. We believe, the present device will be suitable to perform many more experiments to understand the fundamental dynamical features of complex plasmas.
\section*{Acknowledgements}
A.S. is thankful to the Indian National Science Academy (INSA) for their support under the INSA Senior Scientist Fellowship scheme. S. A. would like to acknowledge Minsha Shah for her help in Langmuir probe circuit design and instrumentation.\\\\
\textbf{Data availability statement}\\\\
The data that support the findings of this study are available upon reasonable request from the authors.\\\\
\noindent \textbf{References}
\bibliography{references_arxiv}
\end{document}